\documentclass[twocolumn,aps,floats,prb]{revtex4}
\usepackage{epsfig}
\usepackage{bm}
\usepackage{amsmath}
\usepackage{color}
\usepackage{graphicx}
\newcommand{\mr}[1]{{{\mathrm{#1}}}}

\newcommand{\w}{\omega}
\newcommand{\s}{\sigma}
\newcommand{\aid}{a^{\dagger}_{i}}
\newcommand{\ajd}{a^{\dagger}_{j}}
\newcommand{\ai}{a^{\phantom{\dagger}}_{i}}
\newcommand{\aj}{a^{\phantom{\dagger}}_{j}}

\newcommand{\ad}{a^{\dagger}}
\newcommand{\apd}{a^{\phantom{\dagger}}}
\newcommand{\bd}{b^{\dagger}}
\newcommand{\bpd}{b^{\phantom{\dagger}}}

\begin{document}

\title{Dissipative spin dynamics near a quantum critical point:\\ 
Numerical Renormalization Group and Majorana diagrammatics}

\author{S. Florens}
\author{A. Freyn}
\affiliation{Institut N\'eel, CNRS et Universit\'e Joseph Fourier, 25 avenue
des Martyrs, BP 166, 38042 Grenoble, France}
\author{D. Venturelli}
\affiliation{NEST, Istituto Nanoscienze, CNR and Scuola Normale Superiore, Pisa, Italy}
\affiliation{Institut N\'eel, CNRS et Universit\'e Joseph Fourier, 25 avenue
des Martyrs, BP 166, 38042 Grenoble, France}
\affiliation{International School for Advanced Studies (SISSA), Via Bonomea 265,
I-34136 Trieste, Italy}
\author{R. Narayanan}
\affiliation{Department of Physics, Indian Institute of Technology Madras,
Chennai 600036, India}

\begin{abstract}
We provide an extensive study of the sub-ohmic spin-boson model with
power law density of states $J(\w)\propto \w^s$ (with $0<s\leq1$), focusing on 
the equilibrium dynamics of the three possible spin components, from very 
weak dissipation to the quantum critical regime. 
Two complementary methods, the bosonic Numerical Renormalization Group (NRG) 
and Majorana diagrammatics, are used to explore the physical 
properties in a wide range of parameters. We show that the bosonic self-energy
is the crucial ingredient for the description of critical fluctuations, but
that many-body vertex corrections need to be incorporated as well in order to
obtain quantitative agreement of the diagrammatics with the numerical simulations.
Our results suggest that the out-of-equilibrium dynamics in dissipative models 
beyond the Bloch-Redfield regime should be reconsidered in the long-time limit.
Regarding also the spin-boson Hamiltonian as a toy model of quantum 
criticality, some of the insights gained here may be relevant for field 
theories of electrons coupled to bosons in higher dimensions.
\end{abstract}
\maketitle

\section{Introduction}

Quantum dissipative models, introduced on a phenomenological basis in many areas 
of physics, have proved to be an invaluable tool for studying 
the quantum mechanical friction of an external bosonic environment on a generic 
two-state system.~\cite{Leggett,weiss} While most studies on quantum 
dissipative models investigated questions pertaining to decoherence at 
short time scales, their long-time behavior, in 
particular near possible quantum critical points, is less well understood,
see Refs.~\onlinecite{VojtaReview,LeHurReview,FlorensReview} for recent reviews.
The motivation to investigate such regimes comes from the recent interest 
in zero temperature (quantum) phase transitions,~\cite{Sachdev} where
anomalous low energy properties emerge due to the presence of quantum critical modes.
Despite their apparent simplicity, dissipative impurity models are deeply concerned by
this phenomenon, displaying often non-trivial long-time dynamics.
They can be used to describe quantum criticality not only at the level of
a single magnetic impurity diluted in a correlated system (such as Mott 
insulators~\cite{Shraiman_Giam,Novais,Vojta_Bura} or magnetic metals~\cite{Larkin,Loh}),
but also in bulk materials themselves (e.g.\ in quantum spin glasses,~\cite{Georges}
or heavy fermion compounds~\cite{Sengupta, Si}) via the framework of 
the Dynamical Mean Field Theory.~\cite{dmft} Although we restrict our discussion 
here to a two-level impurity model coupled to a bosonic continuum, it is hard not to mention
the host of fascinating physical phenomena in fermionic models, with the Kondo
problem \cite{Hewson} and its variations being another major example for dissipation (via 
magnetic screening) and quantum critical phenomena (see Ref.~\onlinecite{VojtaReview} 
for a recent theoretical review and Ref.~\onlinecite{RochReview} 
for a discussion of some experimental realizations in quantum dots).

Despite their quantum nature, the phase transitions in dissipative quantum 
impurity models are generally believed to be well described by
an associated long-range classical one-dimensional Ising model.~\cite{Emery}
Indeed the sub-ohmic spin-boson model, in which a single spin $S=1/2$ is subjected
to a magnetic field $\Delta$ perpendicular to the quantization axis, and coupled
(along the quantization axis) to a bosonic continuum with vanishing density of 
states $J(\w)=2 \pi \alpha \w_c^{1-s} \w^s\theta(\w)\theta(\w_c-\w)$, obeys such 
quantum-to-classical mapping in the range $1/2<s<1$, as confirmed by recent direct simulations 
of the quantum model.~\cite{Bulla}
In this case, the quantum critical behavior follows non-trivial {\it classical} exponents of 
an Ising chain with long-range couplings, decaying as $1/(\tau-\tau')^{1+s}$ in the
imaginary time domain ($\tau$ variables), and is described by an interacting fixed 
point.~\cite{Fisher, Kosterlitz}
However, the behavior in the strongly sub-ohmic case, $0<s<1/2$, has been recently
debated. According to the quantum-to-classical mapping, the model should fall
above its upper critical dimension and show mean-field exponents with a
violation of hyperscaling laws. Initial evidence for a different behavior,
namely non-mean-field exponents above the upper critical dimension for the
strongly sub-ohmic spin-boson model, came with surprise from Numerical Renormalization 
Group (NRG) calculations.~\cite{Vojta} Ref.~\onlinecite{Kirchner} also provided
similar puzzling results from Monte Carlo simulations based on a truncation of the long-range 
interaction (which may or may not be in the same universality class as the
spin-boson problem).
At this juncture it is to be noted that these results have appeared recently to be
controversial,~\cite{Rieger,Erratum,Anders,Feshke,Zhang,TongTrunc} and at
present no final consensus has appeared
on the definite nature of this continuous quantum phase transition (QPT).  
Still, there are hints that the NRG displays truncation errors in the long-range
ordered phase~\cite{Erratum,Anders,TongTrunc}, so that the quantum-to-classical mapping may
hold after all.
We note that a discontinuous first order transition, as obtained from variational
methods,~\cite{Variational} is likely an artifact resulting from a
breakdown of this approximation in the quantum critical regime, and that the
localization transition in the spin-boson model is currently believed to be
second order.

In the present paper we will not attempt to answer all these unresolved issues, 
yet we hope to make progress on the quantitative understanding of the spin-boson model 
near its quantum critical point, focusing on the so-called delocalized regime, 
which corresponds to the classically disordered phase. Two complementary tools will be 
developed, namely NRG
calculations~\cite{Wilson,BullaRMP,Vojta,Tong} and Majorana-fermion diagrammatic 
theory,~\cite{Maj1,Maj2} that will be benchmarked against each other, both far 
from and close to the quantum critical point.
Although the Majorana modes appear here as a technical device to reproduce the
spin dynamics of a standard two-level system, the methodology developed here
may be useful too in the emerging field of Majorana qubits.~\cite{Hasan} 
Our work will also aim at an exhaustive study of the physical properties of the
model, and we will examine in detail the zero-temperature {\it equilibrium} spin 
dynamics in all three possible spin directions, which was previously not achieved 
to our knowledge.
In this context, it is also interesting to use the NRG as a testbed for
diagrammatic methods in fermion-boson models, and we will discuss how various
GW-like schemes compare with the numerics. Interestingly, we will show that the
phase transition is driven by a Bose-condensation of the mode mediating the
interaction between Majorana fermions, and that ladder resummation
within the bosonic self-energy is the crucial ingredient to obtain the precise
location of the quantum phase transition. This allows us to establish (at
two-loop order) a phase boundary given by the following critical dissipation 
$\alpha_c = (s/2+s^2/4) |\Delta/\w_c|^{1-s}
+\mathcal{O}(s^3)$, that describes the NRG phase diagram quite accurately for
all values of $\Delta/\w_c<1$ as long as $s\lesssim0.8$.
We believe that the precise understanding achieved here will be important to 
make further progress in the elucidation of quantum critical properties of the model.
It is also worth stressing that the {\it non-equilibrium}
dynamics,~\cite{Leggett} more often considered in the context of
spin-qubits,~\cite{Shnirman} should also be deeply affected at long-times
in case of proximity to a quantum critical points, see e.g.\
Ref.~\onlinecite{Thorwart} for a related study. Standard master equation methods
at the level of Bloch-Redfield approximations (analogous to lowest order
perturbation theory in our equilibrium computations) should fail in this regime,
and ought to be reconsidered seriously, possibly at the light of further
developments of both perturbative techniques~\cite{Whitney,Nesi} and out-of-equilibrium NRG.~\cite{Anders2}

The plan of the paper is as follows. Section~\ref{general} will present the
general properties of the spin-boson model, and its solution at zero-temperature by the NRG.
An improved broadening method~\cite{Freyn} will be used to extract the dynamical
spin susceptibilities (the longitudinal as well as the transverse ones).
Section \ref{Majorana} will present the Majorana-fermion diagrammatic method,
based on perturbation theory in the dissipation strength, focusing on the regime
of very weak dissipation far from the quantum critical point. Then, we will
extend the diagrammatics up to the vicinity of the quantum phase transition
in Section~\ref{QCP}. Leading logarithmic corrections in the form of ladder
diagrams will be obtained in a two-loop calculation of the bosonic self-energy, 
leading to the accurate analytical formula for the phase boundary quoted above. Such renormalization 
effects, taking into account quantitative vertex corrections, will allow to extend the 
Majorana diagrammatics non-perturbatively, and match 
the numerical data both at low (critical modes) and high energy (dissipative features). 
Further directions of research will be proposed as a conclusion, and
several appendices will contain some technical details, such as a
derivation of Shiba's relation for the sub-ohmic spin-boson model, and an
heuristic discussion of the quantum-to-classical mapping via the effective
bosonic action.

\section{Spin-boson model and its solution by the NRG}
\label{general}

\subsection{Hamiltonian and physical aspects}
The (sub)-ohmic spin-boson Hamiltonian involves a three-component half-integer 
spin $\vec{S}= \frac{\hbar}{2} \vec{\sigma}$ (we set $\hbar=1$ in what follows), 
with $\vec{\sigma}$ the three Pauli matrices, and a continuous bath of bosonic oscillators $\aid$:
\begin{equation}
H = \frac{\Delta}{2} \s^x + \frac{\epsilon}{2} \s^z + \frac{\lambda}{2} \s^z 
\sum_i (\aid+\ai) + \sum_i \w_i \aid \ai .
\label{ham}
\end{equation}
$\Delta$ and $\epsilon$ are magnetic fields applied to the spin (respectively
orthogonal and parallel to the quantization axis), and the coupling constant $\lambda$ controls 
the strength of the dissipation to the environment. The bosonic spectrum (which
includes by standard definition the dissipative coupling $\lambda$) is assumed 
to be a power-law up to a sharp high-energy cutoff $\w_c$:
\begin{eqnarray}
\nonumber
J(\w) & \equiv & \sum_i \pi \lambda^2 \delta(\w-\w_i) = \pi \lambda^2
\frac{\w^s}{\w_c^{1+s}}\theta(\w)\theta(\w_c-\w)\\
J(\w) & = & 2 \pi \alpha \w_c^{1-s}
\w^s\theta(\w)\theta(\w_c-\w)
\label{bath}
\end{eqnarray}
thus defining the dimensionless dissipation strength $\alpha=\lambda^2/(2\w_c^2)$, 
an important parameter of the model.
In all what follows, we will consider a vanishing magnetic field parallel to the
bosonic bath $\epsilon=0$, focusing only on the interplay of dissipation and
spin precession, which are driven respectively by the parameters $\alpha$ and $\Delta$.
Temperature will also be set to zero, so that the model (\ref{ham}) displays
two distinct kinds of ground state: i) a delocalized phase, with zero magnetization 
along the $z$ axis, $\big<S^z\big>=0$, in which the ground state is non-degenerate and is
adiabatically connected to a coherent superposition of the two spin up/down
states as the dissipation strength $\alpha$ is increased from zero (keeping a
finite value of the magnetic field $\Delta$);
ii) a localized phase, with a doubly degenerate ground state, where dissipation 
is strong enough to polarize the spin, leading to a non-zero magnetization
$\big<S^z\big>\neq0$, associated to a spontaneous symmetry breaking. We
emphasize that in the present quantum problem, long-range correlations associated 
to the slow decay in time of the bath correlations can be sufficient to generate
long-range order at zero-temperature.
When $0<s\leq1$, a quantum phase transition therefore occurs between the two phases at
a critical value $\alpha_c$ of the dissipation strength. Having in mind a
perturbative expansion in the dissipation, we will focus in this work on the delocalized
phase only up to the quantum critical point, i.e.\ $\alpha<\alpha_c$.
We also note that in {\it both} phases, a finite magnetization along the applied
magnetic field $\Delta$ is always present, $\big<S^x\big>\neq0$.

\begin{figure}
\includegraphics[scale=1.0]{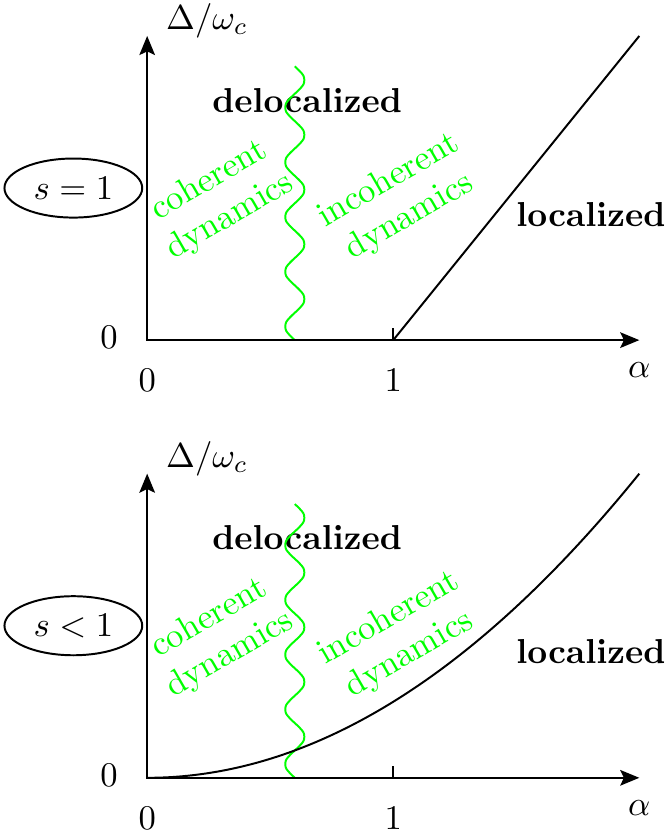}
\vspace{0.1cm}\\
\caption{Sketch of the generic phase diagram of the spin-boson model, in the ohmic ($s=1$) and
sub-ohmic ($0<s<1$) cases respectively, as a function of the dissipation
parameter $\alpha$ and the dimensionless magnetic field $\Delta/\w_c$. 
The super-ohmic ($s>1$) regime shows only the crossover from coherent to incoherent
dynamics, without a localization transition. In all cases, the
coherent/incoherent crossover happens at comparable coupling strength, namely
$\alpha\simeq0.5$.}
\label{PhaseDiagram}
\end{figure}

Several exact results are known for the model~(\ref{ham}), see the reviews in 
Refs.~\onlinecite{Leggett,VojtaReview}.
For $s=1$ (ohmic case), the quantum phase transition is of Kosterlitz-Thouless type, 
and is intimately connected to the ferromagnetic-antiferromagnetic transition in the Kondo model. 
This connection, that can be precisely shown with the use of
bosonization,~\cite{Leggett,Costi} makes it possible to demonstrate
that the critical dissipation strength is exactly $\alpha_c=1$ in the limit $\Delta\ll\omega_c$. 
A particularity of the ohmic case ($s=1$) is that the crossover from coherent to incoherent spin 
dynamics (corresponding respectively to weakly damped and overdamped Rabi
oscillations in the transverse spin autocorrelation functions) occurs {\it before} the onset 
of the QPT, so that the localized phase has always an incoherent character, see 
Fig.~\ref{PhaseDiagram}.
In contrast, the sub-ohmic model ($0<s<1$) can undergo localization before
or after the disappearance of the coherent oscillations, depending on the
precise values of the microscopic parameters,~\cite{Anders2,Thorwart} see
Fig.~\ref{PhaseDiagram}. In particular, in the limit of small magnetic field
$\Delta/\w_c\ll1$, the quantum phase transition occurs at small dissipation
$\alpha\ll1$, so that low-energy dynamics (related to quantum criticality)
and high energy physics (the damping of the Rabi oscillations) are
governed by different energy scales, and both can be treated in principle by
perturbative methods. We will focus in this section on a Numerical
renormalization group (NRG) solution that encompasses both weak ($\alpha\ll1$) 
and non-perturbative dissipation ($\alpha\simeq0.1$) regimes, before making comparison with 
analytical methods in Sections~\ref{Majorana} and~\ref{QCP} for the weak 
dissipation case.

\subsection{Bosonic NRG algorithm}

The implementation of the NRG method follows the standard procedure initially
introduced for fermionic models,~\cite{Wilson,BullaRMP} and recently extended to
bosonic Hamiltonians.~\cite{Tong}
First, the spin-boson model~(\ref{ham}) is rewritten in a continuous form:
\begin{equation}
H=\frac{\Delta}{2}\sigma^x
+\frac{1}{2}\sigma^z \int_0^{\w_c} \!\!\!\!\! d\epsilon \; h(\epsilon)\; 
[a^\dagger_\epsilon +a^{\phantom{\dagger}}_\epsilon] + 
\int_0^{\w_c} \!\!\!\!\! d\epsilon \; \epsilon \; a^\dagger_\epsilon a^{\phantom{\dagger}}_\epsilon 
\end{equation}
with $J(\epsilon)/\pi=[h(\epsilon)]^2$.
The bosonic bath $J(\epsilon)$ is then logarithmically discretized using the
Wilson parameter $\Lambda>1$, first on the highest energy interval near the
cutoff $[\Lambda^{-z}\w_c,\w_c]$, and then iteratively on successive decreasing energy
windows $[\w_{n+1},\w_n]$ with $\w_n=\Lambda^{-n-z+1}\w_c$ (for $n$ strictly positive
integer).
This choice of discretization introduces the so-called $z$-parameter (with $0<z\leq1$) 
that is used to average over $N_z$ different realizations of the Wilson
chain~\cite{Oliveira1} (this amounts to gradual changes of boundary conditions,
allowing to obtain better statistics from the numerical simulations).
Taking $N_z$ different $z$ values, usually uniformly distributed in the $]0,1]$
interval, allows to obtain an order $N_z$ improvement of the spectral resolution at 
finite energy. For very narrow spectral structures in the correlation functions, this
method becomes too expensive and and one lays recourse to improved broadening techniques, see Ref.~\onlinecite{Freyn}
and Sec.~\ref{btrick}.

The bosonic fields are then decomposed in Fourier modes (with $p$ a positive or
negative integer) on each interval $\w_{n+1}<\epsilon<\w_n$ of width
$d_n=(1-\Lambda^{-1})\Lambda^{-n-z}$ for $n>0$ (note that $d_0=1-\Lambda^{-1}$):
\begin{eqnarray}
\label{field}
\ad_\epsilon = \sum_{n,p} \frac{e^{i 2\pi p \epsilon/d_n}}{\sqrt{d_n}}
\, a^\dagger_{n,p}.
\end{eqnarray}
This step is of course exact, and the first NRG approximation consists in neglecting 
all $p\neq0$ modes, keeping only the operator $\ad_n \equiv a^\dagger_{n,0}$ (the full 
Hamiltonian is thus only recovered in the $\Lambda\to1$ limit.~\cite{Bulla})
This leads to the ``star''-Hamiltonian:
\begin{equation}
H_\mathrm{star}=\frac{\Delta}{2}\sigma^x
+\frac{1}{2}\sigma^z \sum_{n=0}^{+\infty} \frac{\gamma_n}{\sqrt{\pi}}
[\ad_n+\apd_n]+ \sum_{n=0}^{+\infty} \xi_n \ad_n \apd_n
\label{star}
\end{equation}
with the ``impurity'' coupling strength (the expressions below apply for $n>0$
only, the $n=0$ value can be obtained analogously)
\begin{equation}
\gamma_n^2=\int_{\w_{n+1}}^{\w_n}\!\!\!\! d\w\, J(\w) =
2\pi\alpha \frac{1-\Lambda^{-(s+1)}}{s+1} \w_c^2 \Lambda^{-(n+z)(s+1)}
\label{gamma}
\end{equation}
and the typical energy $\xi_n$ in each Wilson shell
\begin{equation}
\xi_n = \frac{1}{\gamma_n^2} \int_{\w_{n+1}}^{\w_n}\!\!\!\! d\w\, \w\, J(\w) =
\frac{s+1}{s+2}\frac{1-\Lambda^{-(s+2)}}{1-\Lambda^{-(s+1)}}\w_c
\Lambda^{-n-z}.
\label{xi}
\end{equation}
One then performs an exact mapping onto the so-called chain Hamiltonian:
\begin{eqnarray}
\label{chain}
H_\mathrm{chain}&=&\frac{\Delta}{2}\sigma^x
+\frac{1}{2}\sigma^z \frac{\eta_0}{\pi} [\bd_0+\bpd_0]\\
\nonumber
&&+ \sum_{n=0}^{+\infty} [\epsilon_n \bd_n \bpd_n+t_n(\bd_{n+1} \bpd_n+\bd_n \bpd_{n+1})]
\end{eqnarray}
where $\eta_0=\int d\epsilon J(\epsilon)$ and the on-site energies $\epsilon_n$
and intersite hoppings $t_n$ are determined numerically by a tridiagonalization
of the bosonic part of the star Hamiltonian~(\ref{star}).

The remaining step of the NRG follows Ref.~\onlinecite{Tong}, adding
successively sites into $H_\mathrm{chain}$ starting with $n=0$. The site of order $n$ 
roughly describes the physics for energies of the order $\w_c \Lambda^{-n}$.
This iterative procedure allows to reach exponentially small scales in a linear
effort, an important achievement for the study of Kondo physics~\cite{Wilson} or
impurity quantum critical behavior.~\cite{VojtaReview} It is important to stress
that the success of the NRG relies on the exponential decrease of the chain
parameters $\epsilon_n$ and $t_n$ (related to the analogous behavior of the star
parameters $\xi_n$ and $\gamma_n$), which makes the iterative diagonalization
reliable and stable (see however Ref.~\onlinecite{Freyn2} for the extension of
NRG to other classes of models). Because the size of the Hilbert space needs to remain finite,
truncation constraints need to be implemented.
Typical calculations are usually performed with $N_b=10$ bosonic states on each 
successive bosonic site of the Wilson chain that is added during the
renormalization procedure,~\cite{Tong,Freyn} and $N=160$ kept NRG states (truncation). 
Matrix sizes thus do not exceed $2000\times2000$ during the whole computation, and
we proceed to $N=30$ NRG iterations with $\Lambda=2$ , so that a typical low energy 
scale $\w_\mathrm{min}=\w_c \Lambda^{-N}\simeq 10^{-9}$ can be reached.
The resulting discrete energy spectra at successive NRG iterations are combined 
using the interpolation scheme proposed in Ref.~\onlinecite{Bulla3} (see however
the more rigorous implementation given by full density-matrix NRG 
calculations~\cite{Weichselbaum,Peters}), leading to a set of $z$-dependent many-body 
energy levels $\epsilon_{k,z}$ labelled by quantum number $k$.
The spin-spin correlation function along the $i$-axis ($i=x,y,z$) is defined 
from the equilibrium spin autocorrelation functions in real time and their Fourier transform 
to real frequency:
\begin{eqnarray}
C_i(t)&=&\frac{1}{2} \big<[\sigma_i(t),\sigma_i(0)]\big>\\
C_i(\w)&=&\int \frac{\mathrm{d}\w}{2\pi} \, C_i(t) \cos(\w t).
\label{autocorrelation}
\end{eqnarray}
An alternative way to introduce the spin dynamics is from the imaginary time
spin correlation functions and their Fourier transform onto Matsubara frequency:
\begin{eqnarray}
\label{susceptibilities1}
\chi(\tau)&=&\big<S_i(\tau)S_i(0)\big>\\
\chi(i\nu)&=& \int_0^\beta \!\!d\tau\, e^{i\nu\tau} \chi(\tau)\\
\chi(\w+i0^+)&=&\chi'(\w)+i\chi''(\w)
\end{eqnarray}
where analytic continuation was performed in the last equation in order to obtain the 
retarded spin susceptibility. These susceptibilities are simply related to the spin autocorrelation 
functions~(\ref{autocorrelation}) by the fluctuation-dissipation theorem, which
reads at equilibrium and for zero temperature (we consider the $T=0$ limit in all what follows):
\begin{eqnarray}
\label{susceptibilities2}
C_i(\w)&=& \mathrm{Sign}(\w) \chi_i''(\w).
\end{eqnarray}
We will therefore use equivalently both nomenclatures in the rest of the paper.
There exists an important equation, known as Shiba's relation,~\cite{Costi,Tong}
which connects real and imaginary parts of the spin susceptibility at low
frequency. This exact relation is usually derived for the ohmic ($s=1$) spin-boson model (or 
equivalently for the anisotropic Kondo model) and reads at small frequency:
\begin{equation}
C_z(\w)= 2\pi\alpha|\w| [\chi'_z(0)]^2.
\end{equation}
We will prove in appendix~\ref{shiba} that this result can be generalized to
the sub-ohmic case ($0<s<1$) in the following form
\begin{equation}
C_z(\w)= J(|\w|) [\chi'_z(0)]^2
\label{shibasubohmic}
\end{equation}
valid in the small $\w$ limit.
An useful byproduct of Shiba's relation is that, at a magnetic ordering of the 
spin (corresponding to the quantum critical point), the susceptibility
$\chi'_z(0)$ diverges, so that the low-energy power law behavior 
$C_z(\w)\propto J(|\w|)\propto |\w|^s$ obeyed in the whole delocalized phase should 
turn into a different (and diverging) power law.

A last important technical step is that all these correlation functions can be 
computed at zero temperature from the raw NRG data using Lehmann's decomposition rule:
\begin{equation}
C_i(\w )=\frac{1}{2 N_z}\sum_{k,z} | \langle 0,z | \sigma_i | k,z \rangle |^2
\delta \left( |\w|+\epsilon_{0,z}-\epsilon_{k,z} \right)
\label{Lehmann}
\end{equation}
where $\epsilon_{0,z}$ is the ground state energy.
This leads to a superposition of sharp $\delta$-peaks which need to be
broadened. This sensitive issue is examined now.

\subsection{Optimized broadening method}
\label{btrick}

\begin{figure}[ht]
\includegraphics[width=8.5cm]{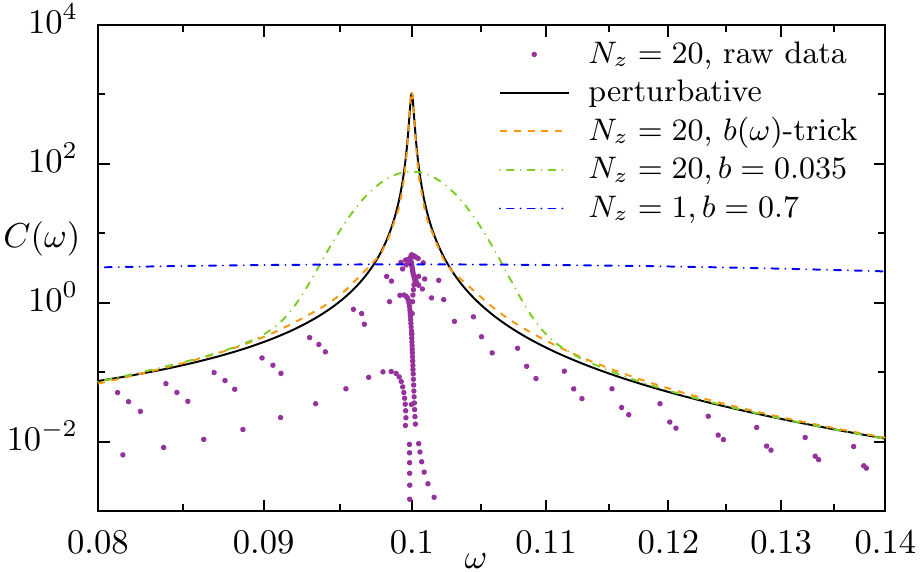}
\caption{(Color online) The transverse (along the bath) spin susceptibility $C_z(\w)$ 
of the sub-ohmic spin-boson model at
$s=0.1$, $\Delta=0.1\w_c$ and $\alpha=0.000125$. Raw data
$| \langle 0,z | \sigma_z | k,z \rangle |^2/[2N_z
(\epsilon_{k,z}-\epsilon_{0,z})]$ are
given as circles for $N_z=20$ combined NRG calculations, solid line is the
perturbative result, and the three dashed lines are the various NRG broadenings 
discussed in the text. All frequency-dependent plots are given in units of
$\w_c=1$.}
\label{fig:raw}
\end{figure}

A single NRG calculation (for a given value of $z$) usually provides a set 
of energy levels which come by packets located around each Wilson shell.
The energy resolution at scale $\w_n=\Lambda^{-n-z}$ is thus usually 
thought to be of the order $d_n= \w_n-\w_{n+1}=(1-\Lambda^{-1})
\w_n$, and degrades at higher energy. 
In order to generate smooth NRG spectra, the delta-peaks in the Lehmann formula~(\ref{Lehmann}) 
are therefore usually broadened~\cite{BullaRMP} at energy $\w_n$ {\it on a scale of
the same order} (the broadening scale is typically $b \w_n$, with $b\simeq0.7$) using the 
substitution:
\begin{equation}
\delta \left( |\w|-\w_n\right) \rightarrow
\frac{e^{-b^2/4}}{\w_n b \sqrt{\pi}}
e^{-[\log(|\w|/\w_n)^2/b]^2}.
\label{eq:broaden}
\end{equation}
As a result, spectral features at a given energy $\w$ that are sharper than $\w$
itself cannot be well resolved, and will come out overbroadened.
Some relative degree of improvement can be achieved by combining $N_z$ NRG runs with
different values of the $z$ parameter ($z$-averaging procedure, or
$z$-trick~\cite{Oliveira1}).
Although the mathematical justification for this procedure is still unclear, in
the optimal case the resulting energy packets turn out to be uniformly distributed if
the set of $z$-values is also chosen uniformly, and this allows a decrease of
the broadening parameter to $b\simeq0.7/N_z$. In the case of very sharp features,
this method requires a large number of NRG runs, and will be prohibitive. It
may also become problematic in regions where the $z$-averaging does not provide
a uniform distribution of states (this may occur from the accidental disappearance
of some NRG states), in which case uncontrolled oscillations of period $\Lambda$ can 
be generated (situation of underbroadening).
These problems have led authors to speculate that quantitative NRG spectral
functions can be extracted only in the continuum limit,~\cite{Zitko,Oliveira2,Weichselbaum2}
either with $\Lambda\rightarrow1$ or $N_z\rightarrow+\infty$ (only the former is
mathematically sound, but the NRG algorithm cannot be managed anymore as scale
separation breaks down).
Both limitations of the $z$-averaging can however be lifted thanks to a simple
remark made by two of us in a previous publication:~\cite{Freyn} for a single NRG 
run, the levels within a given logarithmic energy window are not uniformly distributed, 
but rather tend to bunch together close to sharp resonances. This effect becomes
quite obvious when the width of such resonance becomes extremely small, see 
Fig.~\ref{fig:raw}, but is always present in all NRG calculations.
Figure ~\ref{fig:raw} shows indeed the raw NRG spectra for one of the
spin susceptibilities of the spin-boson model, computed with $N_z=20$ interleaved
$z$-averaged NRG runs (dots) and smoothed using two standard NRG broadenings 
(dash-dotted curves) with respective parameters $b=0.7$ and $b=0.7/N_z=0.035$.
The largest broadening clearly fails to produce the expected peak
(as a testbed, an accurate analytical expression is provided by the full line,
see Sec.~\ref{Majorana} for further details), while the smallest
broadening signals the peak, but overestimates its width by more than
one order of magnitude! Large scale $z$-averaging with $N_z\simeq10^4$ would
probably allow to resolve the correct spectral structure, but is unreasonably expensive.
The key observation~\cite{Freyn} is that the NRG eigenvalues do not always come in 
packets uniformly distributed within the Wilson shell, but on the contrary tend 
to cluster close to resonances, as clearly seen on the raw data (dots) in 
Fig.~\ref{fig:raw}. This provides the missing information which is needed in
order to achieve a good spectral resolution with a limited numerical cost. 
The main strategy to be used is that {\it the broadening parameter $b$
must become energy-dependent}, which we previously called the
``b-trick''.~\cite{Freyn} However, the generic implementation of this simple idea
is not fully clarified: we present here a simple scheme that seems to be quite
efficient, but we emphasize that a more generic and robust method is still to be
found. The required algorithm must indeed adapt the broadening parameter $b(\w)$ 
in Eq.~(\ref{eq:broaden}) to the frequency dependence of the density of highest weight NRG peaks. 
Our approach, which draws inspiration from the NRG logarithmic discretization, 
is to extract $b(\w)$ from the logarithmic derivative of the integrated spectrum 
up to frequency $\w$:
\begin{equation}
b(\w) = \frac{b_0}{2}\left(\left[q+\frac{\mr{d}\log \int_0^\w
C}{\mr{d}\log\w}\right]^{-1}\!\!\!\!\!\!+\!\left[q+\frac{\mr{d}\log \int_{+\infty}^\w
C}{\mr{d}\log\w}\right]^{-1}\right)
\label{eq:newbroaden}
\end{equation}
where $q\simeq1$ is a regularization parameter whose precise value does not
matter much to the final result, and $b_0$ provides the typical broadening at 
low and high frequencies (far from the atomic resonances). Note that we have
used here two different frequency sweeps, one from $\w=0$ and one from $\w=+\infty$, 
in order to treat on an equal basis low and high frequency tails (this procedure is 
not satisfactory in cases where several resonances are present).
Because the actual NRG data is fully discrete, see Eq.~(\ref{Lehmann}), we compute $b(\w)$ 
{\it recursively} using Eq.~(\ref{eq:newbroaden}) on the broadened NRG spectra. 
This procedure converges after few iterations to the results displayed on Fig.~\ref{fig:raw}.
It is clearly remarkable that such a good agreement can be reached with so
little numerical effort, demonstrating that the raw NRG data encode much more
information than previously thought (see Ref.~\onlinecite{Freyn} for further
details).
One important advantage of the b-trick is that $b(\w)$ takes very small values
near resonances, greatly enhancing the resolution, while keeping large values
away from the peak, thus avoiding the usual underbroadened NRG oscillations.
The main tuning parameter in our method is the low-frequency broadening $b_0$ 
in Eq.~(\ref{eq:newbroaden}), which is determined by a simple convergence method.
We finally emphasize that some degree of z-averaging is always needed in order
to generate smooth spectra with the b-trick, but in our experience z-averaging 
alone is never sufficient to obtain accurate results near sharp finite-frequency
resonances.

\subsection{NRG results for the dynamical susceptibilities in the three spin directions}
\label{dynamics}

We present here the NRG data for a fixed value of the magnetic field
$\Delta/\w_c=0.1$, varying the dissipation from values $\alpha\ll\alpha_c$
(far from the quantum critical point) up to $\alpha_c$. We choose three
different values of the bath exponent $s=0.1,0.5,1.0$, which span the
complete generality of the model. We remark that the case $s=0$ is special, 
as the spin is localized for all $\alpha>0$ at zero temperature~\cite{Anders2},
and will not be considered here, as we aim at understanding the delocalized
phase of the model only. We will also not examine the situation of $s>1$, which 
shows  for weak dissipation similar features as the $s=1$ case, and does not present
a phase transition at larger coupling. We stress already that depending whether
$\alpha_c\ll1$ or not, the transition will or won't take place at weak coupling,
and this can lead to different physical pictures.

\begin{figure}[ht]
\includegraphics[width=7.0cm]{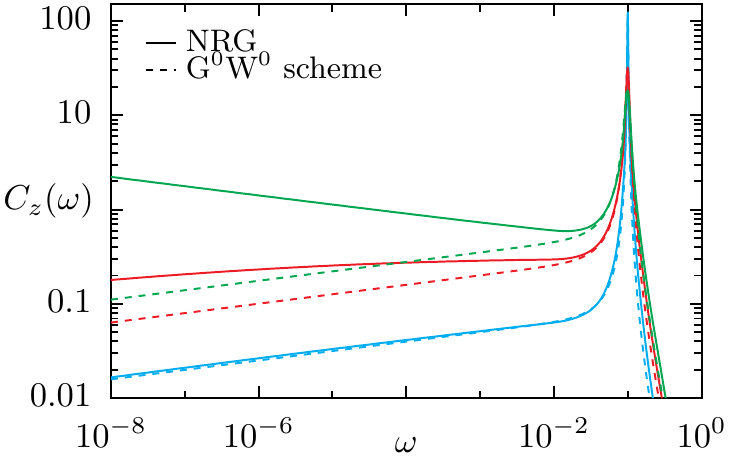}
\includegraphics[width=7.0cm]{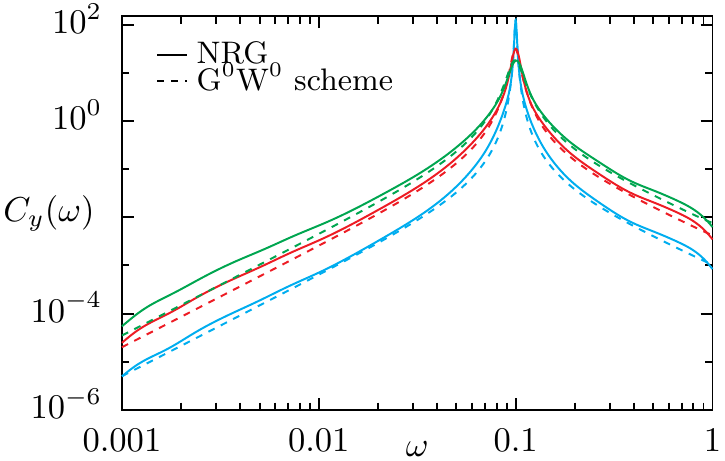}
\includegraphics[width=7.0cm]{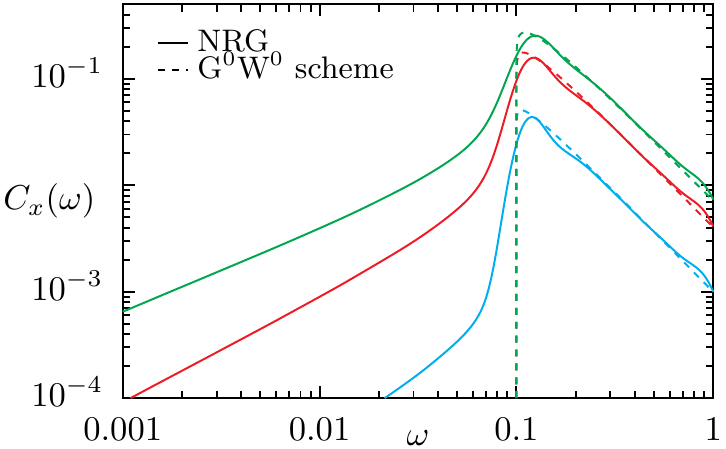}
\caption{Spin susceptibilities $C_z(\w)$, $C_y(\w)$ and $C_x(\w)$
(top to bottom panels) for $\Delta/\w_c=0.1$, $s=0.1$ and increasing
dissipation strength $\alpha=0.001,0.004,0.007$ (bottom to top). The critical point is located
at $\alpha_c\simeq0.007$. The NRG plots were generated with $N_z=20$ interleaved
z-averages and used the b-trick discussed in Sec.~\ref{btrick}, while the
perturbative expressions (G$^0$W$^0$ scheme) are given in
Eqs.~(\ref{SigmaxPert}-\ref{SigmayPert}).
Note that the upper panel has a different range in $\w$ than the two other panels.}
\label{NRGChi_s0.1}
\end{figure}

Let us consider first the case $s=0.1$ and the spin susceptibilities along
each spin direction ($i=x,y,z$) shown in Fig.~\ref{NRGChi_s0.1} (the plot
also includes a comparison to lowest order pertubation theory, introduced in
Section~\ref{lowest}).
The transverse (orthogonal to the field and parallel to the bath) spin
susceptibility $C_z(\w)$, already
considered,~\cite{Anders2,Freyn} shows two interesting features. First, the
resonance at $\w=\Delta$ broadened by the bath is always sharply defined, because
here $\alpha_c\simeq0.007$ is so small that one remains in the perturbative
regime for all $\alpha<\alpha_c$ values regarding this high energy features. 
This is consistent with the near perfect agreement with the perturbative result
in this range of frequencies, and weakly damped oscillation should occur in
the real-time spin dynamics.
The low-energy part of this spectral function is yet more interesting. For
$\alpha\ll\alpha_c$, a slow decay with $C_z(\w)\propto |\w|^s$ is
observed, also consistent with the perturbative result. Increasing $\alpha$,
a crossover at a scale $\w^\star$ is seen between the $|\w|^s$ behavior
at $\w\ll\w^\star$ and a different power law $|\w|^{-s}$ at
$\w^\star\ll\w\ll\Delta$. At the quantum critical point $\alpha=\alpha_c$, 
$\w^\star$ vanishes and a complete $|\w|^{-s}$ divergence is realized throughout. 
Clearly, the perturbative expression fails to reproduce this behavior. 
We stress here that even values relatively far from the critical point, such as 
$\alpha=0.004$, show large quantitative deviation from lowest order perturbation theory 
at low frequency, despite the good agreement on the coherent peak. This shows that the
range of validity of perturbative methods (such as Bloch-Redfield), although possibly
accurate at short time scales, is very limited in the long time limit.
Clearly the high energy dissipation mechanism does not care for the 
complex low-energy behavior of the spin dynamics, even at the quantum critical
point.~\cite{Tong,Anders}
Turning to the second transverse susceptibility $C_y(\w)$ (orthogonal to both the 
magnetic field and the spin-bath term), we observe similar behavior as in $C_z(\w)$
for the resonant peak, and good agreement with perturbation theory. The
low energy part of the spectrum is however much less dramatic, because critical
modes, related to fluctuations driven by the bath, pertain mainly to the $z$-component 
of the spin. What is actually going on is a very mild crossover from $\w^{2+s}$ at 
$\w\ll\w^\star$ to $\w^{2-s}$ at $\w^\star\ll\w\ll\Delta$ (the latter behavior
includes the whole range of low frequencies at the critical point), due to the small value 
of $s=0.1$. Agreement with perturbation theory appears thus much better, although small 
deviations, due to the tiny changes of the power law exponent, can be seen at low frequency. 
These precise (and exact) values for the various power laws will be demonstrated in 
Section~\ref{Majorana}.
Finally, the longitudinal (parallel to the bath) spin susceptibility $C_x(\w)$ 
presents quite different features. Clearly in the absence of dissipation ($\alpha=0$), 
a $\delta(\w)$ peak at zero frequency occurs, in contrast to the $\delta(\w-\Delta)$ peak at 
the magnetic field frequency displayed by both $C_y(\w)$ and $C_z(\w)$. 
Thus no sharp resonance is generated at small dissipation, and rather a broad 
shoulder emerges, quite reminiscent of the T-matrix of the Kondo model in a magnetic
field.~\cite{Paaske} This structure is well accounted for by the perturbative
calculation for $\Delta<\w$, however, stark discrepancy is seen at low
frequency for $|\w|<\Delta$: the NRG data present a tail of low energy modes,
while perturbation theory provides a gap in this range. This new feature given by the 
NRG will be elucidated in Section~\ref{multiparticle}, in connection with multiparticle
effects in the diagrammatics, that are neglected in the current lowest order
calculation.

\begin{figure}[ht]
\includegraphics[width=7.0cm]{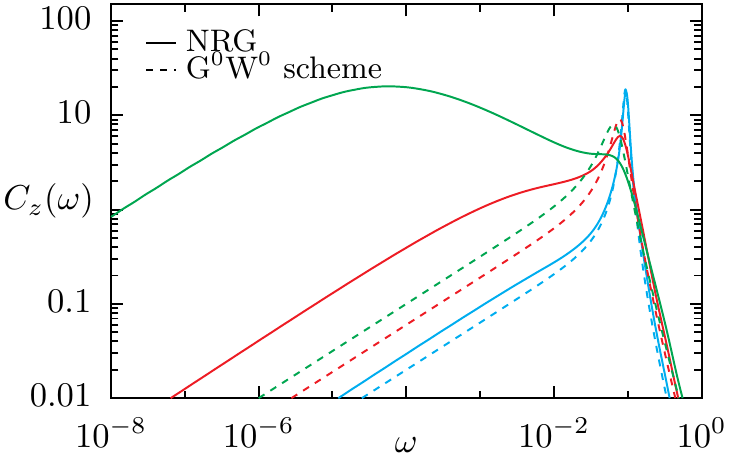}
\includegraphics[width=7.0cm]{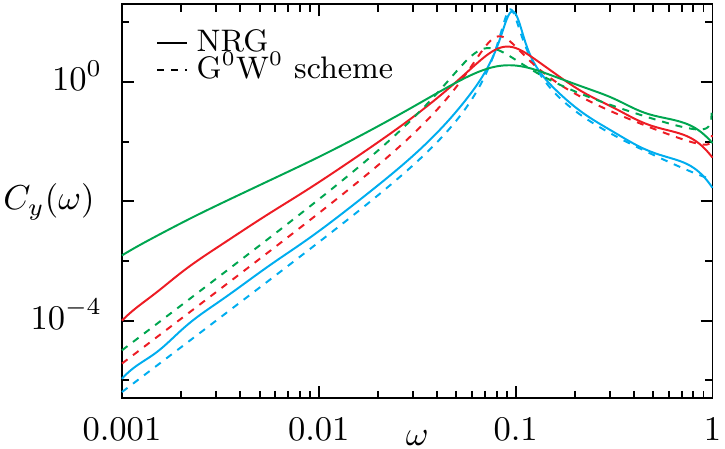}
\includegraphics[width=7.0cm]{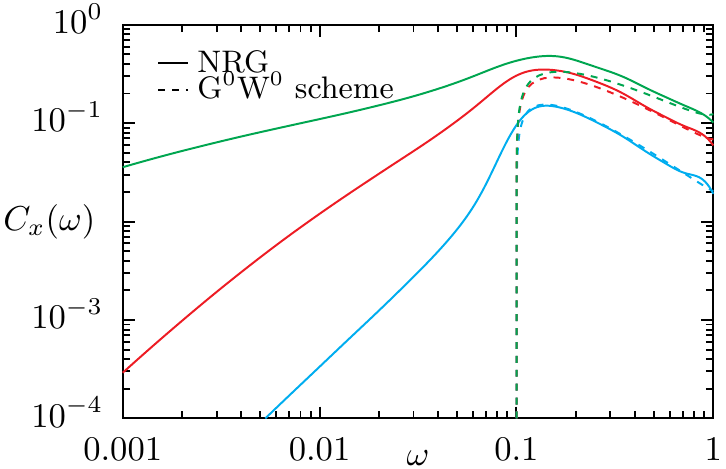}
\caption{Spin susceptibilities $C_z(\w)$, $C_y(\w)$ and $C_x(\w)$
(top to bottom panels) for $\Delta/\w_c=0.1$, $s=0.5$ and increasing
dissipation strength $\alpha=0.02,0.06,0.1$ (bottom to top). The critical point is located
at $\alpha_c\simeq0.105$. The NRG plots were generated with $N_z=10$
interleaved z-averages and used the b-trick discussed in Sec.~\ref{btrick},
while the perturbative expressions (G$^0$W$^0$ scheme) are given in
Eqs.~(\ref{SigmaxPert}-\ref{SigmayPert}).
Note that the upper panel has a different range in $\w$ than the two other panels.}
\label{NRGChi_s0.5}
\end{figure}

We now turn to the intermediate value of $s=0.5$, giving rise to a somewhat larger critical
dissipation strength $\alpha_c\simeq0.105$, see Fig.~\ref{NRGChi_s0.5}.
In that case, the smallest chosen value of $\alpha=0.02$ still lies in the
perturbative regime, so that again near perfect agreement with perturbation 
theory can be checked. However, the approach to the quantum critical point shows
larger quantitative deviations, both at high and low energy. Clearly,
the coherent peak is more strongly smeared in the NRG for the larger shown values
$\alpha=0.06,0.1$ compared to perturbation theory, due to the now relatively
important magnitude of $\alpha$. Nevertheless, the peak structure survives up to
the critical point, so that moderately damped oscillations should show up
in the time domain, even close to $\alpha_c$. Regarding the low energy part of 
the spectrum, the same crossover from $\w^s$ to $\w^{-s}$ behavior (for $\w\ll\w^\star$ 
and $\w^\star\ll\w\ll\Delta$ respectively) is seen in $C_z(\w)$, now more clearly defined
due to the larger value of $s=0.5$. Again perturbation theory fails in
this regime, and this is now more easily seen on $C_y(\w)$ as
well, which presents respectively a crossover from $\w^{2+s}$ to $\w^{2-s}$ in
the frequency dependency. We note also a similar enhancement of the low energy 
tails for the last component $C_x(\w)$ below the shoulder at $\w<\Delta$.

\begin{figure}[ht]
\includegraphics[width=7.0cm]{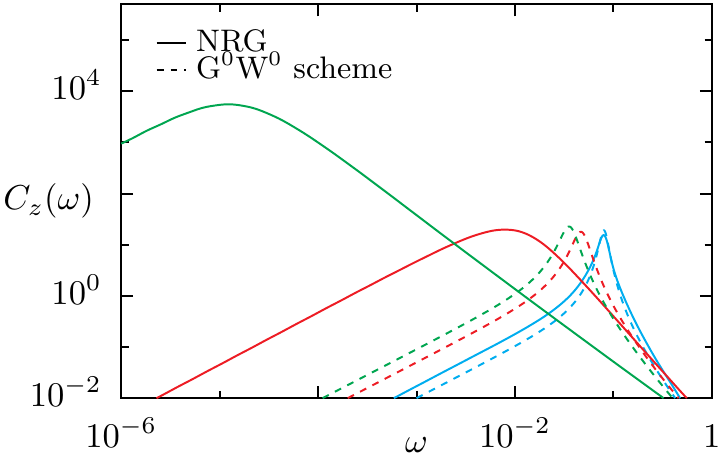}
\includegraphics[width=7.0cm]{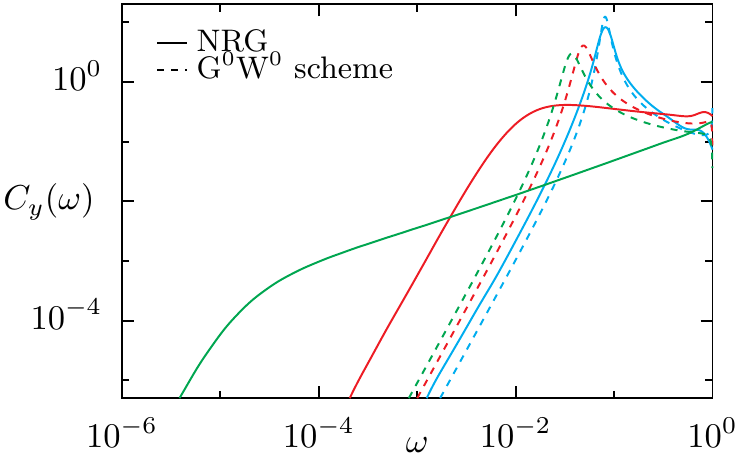}
\includegraphics[width=7.0cm]{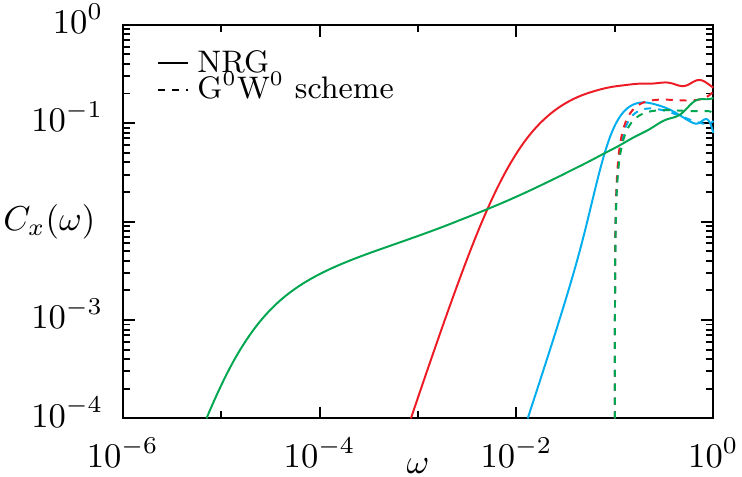}
\caption{Spin susceptibilities $C_z(\w)$, $C_y(\w)$ and $C_x(\w)$
(top to bottom panels) for $\Delta/\w_c=0.1$, $s=1$ and increasing
dissipation strength $\alpha=0.1,0.5,0.9$ (bottom to top). The critical point is located
at $\alpha_c\simeq1$. The NRG plots were generated with $N_z=3$
interleaved z-averages and used the b-trick discussed in Sec.~\ref{btrick},
while the perturbative expressions (G$^0$W$^0$ scheme) are given in
Eqs.~(\ref{SigmaxPert}-\ref{SigmayPert}).}
\label{NRGChi_s1.0}
\end{figure}

We finally consider the so-called ohmic case $s=1$, for which $\alpha_c\simeq1$,
clearly out of the perturbative range, see Fig.~\ref{NRGChi_s1.0}.
Again, small values of $\alpha=0.1$ present well-defined resonance and
reasonable agreement with perturbation theory, although some deviations are
visible. 
However, all values of $\alpha\gtrsim0.5$ show a complete disappearance of the 
coherent peak at $\w=\Delta$, leaving only the crossover scale $\w^\star$, which 
separates regimes where $C_z(\w)\propto\w$ for $\w\ll\w^\star$ to $C_z(\w)\propto\w^{-1}$ 
(up to logarithmic corrections) for $\w^\star\ll\w\ll\w_c$. In that case, oscillations 
are completely overdamped, and coherence is lost already well before the quantum 
critical point $\alpha_c=1$ where the spin becomes logarithmically free (this point 
corresponds precisely to the ferromagnetic scaling limit in the associated Kondo model). 
Again, $C_y(\w)$ shows a similar behavior as observed for smaller $s$ values, with a crossover 
from $\w^{2+s}=\w^3$ to $\w^{2-s}=\w$ power laws for $\w\ll\w^\star$ and $\w^\star\ll\w\ll\Delta$ respectively.
Finally $C_z(\w)$ presents also a similar crossover in the low-energy tails, and
a broad shoulder at the magnetic field value (for weak dissipation).

We now turn to the analytical description of this physics, and show that both qualitative 
and quantitative aspects of the numerical data can be well understood on the
basis of the Majorana-fermion diagrammatic method.

\section{Majorana diagrammatics}
\label{Majorana}

\subsection{Spin representation and correlation functions}

\begin{figure}
\includegraphics[scale=0.90]{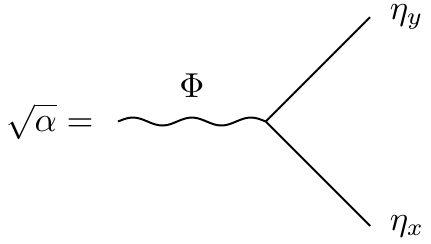}
\vspace{0.1cm}\\
\caption{Bare vertex of the spin-boson model in the Majorana language.}
\label{vertex}
\end{figure}

The idea we henceforth develop is to use a Majorana description~\cite{Maj1,Maj2}
for the impurity quantum spin $\vec{S}=\frac{1}{2}\vec{\sigma}$, in order to capture 
quantitatively the spin fluctuations in the whole delocalized phase up to the
quantum critical point using a weak dissipation expansion (understanding the quantum 
phase transition coming in from the localized phase is a more delicate issue, that we 
will not attempt here).
Technically, our methodology involves the introduction of a triplet of real fermions 
$\eta_j$ (with $j=x,y,z$) that satisfy the anticommutation relations 
$\{\eta_j,\eta_k\} = \delta_{jk}$, leading to the faithful representation 
$S_j = -(i/2)\sum_{kl} \epsilon_{jkl}\eta_k\eta_l$.
We emphasize that Majorana qubits~\cite{Hasan} involve a doublet of real 
fermions, while the mathematical decomposition of a two-level qubits requires 
a triplet of real fermions.
The main motivation behind considering the Majorana representation instead of
more standard ones (Abrikosov or Popov fermions, Holstein-Primakoff or Schwinger bosons) 
is that one can alternatively write $\vec{S} = \chi \vec{\eta}$, where $\chi=-2i\eta_x\eta_y\eta_z$ 
is a fermionic operator which {\it commutes} with the Hamiltonian~(\ref{ham}). This implies
the crucial property that the spin susceptibilities Eq.~(\ref{autocorrelation}) 
are obtained as simple identities involving standard {\it one-particle} Majorana Green's 
function (here at zero temperature):
\begin{equation}
C_i(\w) = -\frac{1}{\pi} \mathcal{I}m \, G_i(\w),
\label{special}
\end{equation}
where $G_i(\w)$ is the retarded real-frequency Majorana-fermion Green's
function, obtained from the Fourier transformation of the time-ordered Green's
function $G_i(t)=-i\theta(t)\big<\{\eta_i(t),\eta_i(0)\}\big>$.
This key relation between $C_i$ and $G_i$ is in great contrast to all 
other spin representations where the spin correlation functions are given in 
terms of {\it four-fermion} correlators, making the diagrammatics much more cumbersome.
Therefore, thanks to Eq.~(\ref{special}), both the dissipative features at 
high energy and the low-energy quantum dynamics close to the quantum critical 
point can be simply encoded via Dyson equation by a Majorana-fermion self-energy. 
To obtain these important relations, we need to reexpress the spin-boson model
Eq.~(\ref{ham}) in the Majorana language:
\begin{equation}
H = -i \Delta \eta_y \eta_z -i\lambda \eta_x \eta_y \sum_j (\ajd+\aj)
+ \sum_j \w_i \ajd \aj .
\label{ham2}
\end{equation}
The ``diagonal'' part given by the magnetic field is readily diagonalized into
``free'' (or bare) Majorana Green's functions:
\begin{eqnarray}
\label{barex}
G_x^{0}(\w)&=&\frac{1}{\w+i0^+}\\
\label{barey}
G_y^{0}(\w)&=&\frac{\w}{(\w+i0^+)^2-\Delta^2}\\
\label{barez}
G_z^{0}(\w)&=&\frac{\w}{(\w+i0^+)^2-\Delta^2}
\end{eqnarray}
which are readily interpreted by a spin precession in the $y-z$ plane around 
the magnetic field pointing in the $x$ direction.
Interaction between the Majorana fermions is generated by the coupling to the
bosonic bath, which can be expressed as a bare vertex involving the $\eta_x$ and
$\eta_y$ field, see Fig.~\ref{vertex}.
This coupling results in three Majorana self-energies $\Sigma_i(\w)$, but clearly
the $z$-component vanishes altogether, $\Sigma_z(\w)=0$, due to the form of the
interaction. Inverting the Majorana propagator (written in matrix form):
\begin{equation}
\hat{G} =
\left[
\begin{array}{ccc} 
\w-\Sigma_x(\w) & 0 & 0 \\
0& \w-\Sigma_y(\w) & i\Delta \\
0& -i\Delta & \w 
\end{array}
\right]^{-1}
\end{equation}
provides Dyson's equation for the full retarded Majorana Green's functions that 
encode the complete spin dynamics:
\begin{eqnarray}
\label{fullx}
G_x(\w)&=&\frac{1}{\w-\Sigma_x(\w)}\\
\label{fully}
G_y(\w)&=&\frac{\w}{\w[\w-\Sigma_y(\w)]-\Delta^2}\\
\label{fullz}
G_z(\w)&=&\frac{\w-\Sigma_y(\w)}{\w[\w-\Sigma_y(\w)]-\Delta^2}
\end{eqnarray}
The bare Green's functions~(\ref{barex}-\ref{barez}) are correctly reproduced in
the zero-dissipation limit, replacing $\Sigma_i(\w)$ by a vanishing imaginary part
$-i0^+$.

Because the interaction term in the Hamiltonian Eq.~(\ref{ham2}) involves a
particular combination of the bosonic modes, it is useful to define the full bosonic 
propagator $G_\Phi(t)= -i\theta(t)\left<[\Phi(t),\Phi(0)]\right>$ of the ``local'' field 
$\Phi\equiv -i \frac{\lambda}{\sqrt{\alpha}} \sum_j (\ajd+\aj)$. This allows us
to introduce a bosonic self-energy $\Sigma_\Phi(\w)$ for the full Green's
function \begin{equation}
\label{fullPhi}
G_\Phi(\w) = \frac{G_\Phi^0(\w)}{1-G_\Phi^0(\w)\Sigma_\Phi(\w)}
\end{equation}
in terms of the bare bosonic propagator
\begin{equation}
G_\Phi^0(\w) = \frac{1}{\alpha} \sum_j \left[ \frac{\lambda^2}{\w+i0^+ -\w_j} 
-\frac{\lambda^2}{\w+i0^+ +\w_j}\right].
\label{GPhibare}
\end{equation}
Using the convention chosen to normalize $\Phi$, the bare vertex takes simply the 
value $\sqrt{\alpha}$ (see Fig.~\ref{vertex}), and the bare bosonic Green's function 
can be expressed through the bosonic spectral density $J(\w)$ given by Eq.~(\ref{bath}):
\begin{eqnarray}
\nonumber
G_\Phi^0(\w) & = & -\frac{1}{\alpha} \int \frac{\mathrm{d} \epsilon}{\pi}
\frac{J(|\epsilon|) \, \mathrm{Sign}(\epsilon)}{\w+i0^+ -\epsilon} \\
& = & - 2 \w_c^{1-s} \int_{-\w_c}^{\w_c} \mathrm{d} \epsilon \,
\frac{|\epsilon|^s \, \mathrm{Sign}(\epsilon)}{\w+i0^+ -\epsilon} 
\label{GPhiBareAlt}
\end{eqnarray} 
a quantity which is indeed independent of the dissipation strength $\alpha$, 
now encoded (by convention) only in the bare vertex. We thus find an explicit
expression for the imaginary part of the bosonic Green's function:
\begin{equation}
\label{ImGPhi}
\mathcal{I}m G_\Phi^0(\w) = 2 \pi \w_c \mathrm{Sign}(\w)
\left|\frac{\w}{\w_c}\right|^s \theta(\w_c^2-\w^2).
\end{equation} 
The real part of the integral~(\ref{GPhiBareAlt}) can be computed analytically
from Kramers-Kronig's relation only in the low-energy limit $|\w|\ll\w_c$:
\begin{equation}
\mathcal{R}e G_\Phi^0(\w) = \frac{4\w_c}{s} -
2 \pi\w_c \mathrm{cotan}\left(\frac{\pi s}{2}\right)
\left|\frac{\w}{\w_c}\right|^s
\label{ReGPhi}
\end{equation} 
although the complete formula~(\ref{GPhiBareAlt}) will be used for later
numerical computations. The low-energy expression~(\ref{ReGPhi})
breaks down for $s=1$, where additional logarithmic corrections arise
(hinting at the known connection to the Kondo model~\cite{Leggett}).

\begin{figure}
\includegraphics[scale=0.90]{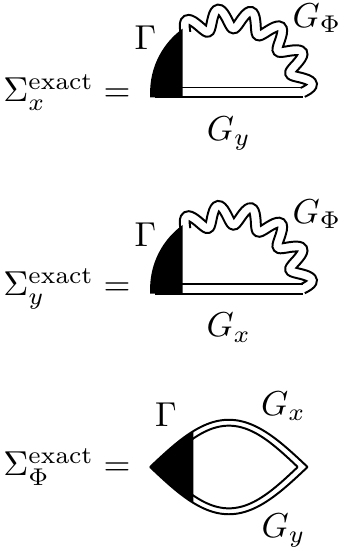}
\vspace{0.1cm}\\
\caption{Exact formal diagrammatic expression for the fermionic and bosonic
self-energies in terms of the full (exact) fermionic and bosonic Green's functions 
($G_i$ and $G_\Phi$ respectively) and the full (exact) three-particle vertex $\Gamma$.}
\label{SigmaExact}
\end{figure}

The exact resummation of perturbation theory requires the accurate knowledge of the full fermionic
and bosonic self-energies, as well as the full three-particle vertex $\Gamma$
(a quantity that depends on three independent external frequencies).
The general diagrammatics is given formally in Fig.~\ref{SigmaExact}.
We note that replacement of the bare propagators by the full ones without considering
together the vertex corrections must be done in general with care, and can lead to
serious pathologies, see Ref.~\onlinecite{Vignale}. An exception to this rule is
given in Sec.~\ref{multiparticle}, and the precise role of the interaction
vertex will be elucidated in Sec.~\ref{SecVertex}.

\subsection{Lowest order perturbative self-energies in the Majorana diagrammatics}
\label{lowest}

\begin{figure}
\includegraphics[scale=0.90]{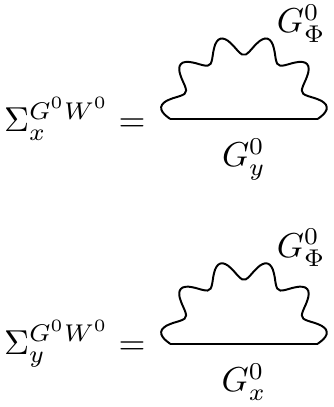}
\vspace{0.1cm}\\
\caption{Lowest order perturbative Majorana self-energies, obtained from bare fermion and
bare bosonic propagators ($G^0W^0$ scheme).}
\label{G0W0}
\end{figure}

For very weak dissipation $\alpha\ll1$ and far from the quantum critical point 
$\alpha\ll\alpha_c$, the spin dynamics should in principle be well described by the 
lowest order perturbative Majorana self-energies (we will see in Sec.~\ref{multiparticle} 
that this naive statement is not completely correct in some specific frequency range). 
These self-energies are obtained at order $\alpha$ by a single boson exchange, see Fig.~\ref{G0W0}, 
where both fermionic and bosonic lines are given by the bare propagators 
Eqs.~(\ref{barex}-\ref{barez}) and Eq.~(\ref{GPhiBareAlt}), respectively. 
In terms of resummation of diagrams, this level of approximation corresponds to the 
so-called $G^0W^0$ scheme for the generic fermion-boson models arising in the fields of 
heavy fermions~\cite{Hertz} or strongly interacting electron liquids~\cite{Vignale} 
($G^0$ denotes the bare fermionic propagator and $W^0\equiv G_\Phi^0$ the bare bosonic one). 
It is alternatively called the (non-self-consistent) Born approximation in the context of disordered 
systems,~\cite{DiCastro} and it roughly corresponds to the weak-coupling Bloch-Redfield 
approximation for the real-time non-equilibrium dynamics of the spin-boson
model.~\cite{Shnirman} Note that the present Majorana diagrammatics is
formulated for the equilibrium dynamics only. Extension to temporal dynamics under a
sudden switching (of the bath or transverse field) could be incorporated in
our formalism by using two-times Keldysh Majorana Green's functions, which lies
however beyond the scope of the present work.
Standard evaluation of the diagrams provides the following imaginary part of 
the Majorana self-energies at zero temperature:
\begin{eqnarray}
\label{SxG0W0}
\mathcal{I}m \Sigma_x^{G^0W^0}\!(\w) & = & \alpha \int_0^\w\!\!
\frac{\mathrm{d}\epsilon}{\pi} \mathcal{I}m G_y^0(\epsilon) \mathcal{I}m
G_\Phi^0(\w-\epsilon) \\
\label{SyG0W0}
\mathcal{I}m \Sigma_y^{G^0W^0}\!(\w) & = &\alpha \int_0^\w\!\!
\frac{\mathrm{d}\epsilon}{\pi} \mathcal{I}m G_x^0(\epsilon) \mathcal{I}m
G_\Phi^0(\w-\epsilon)
\end{eqnarray}
Using the bare propagators~(\ref{barex}-\ref{barez}) and (\ref{GPhibare}),
one readily obtains:
\begin{eqnarray}
\nonumber
\mathcal{I}m \Sigma_x^{G^0W^0}(\w) & = & -\pi \alpha \w_c
\theta[\w_c^2-(|\w|-\Delta)^2] \theta(\w^2-\Delta^2) \\
\label{SigmaxPert}
&& \times \frac{||\w|-\Delta|^s}{\w_c^s}\\
\mathcal{I}m \Sigma_y^{G^0W^0}(\w) & = & -\pi \alpha \w_c
\theta(\w_c^2-\w^2) \frac{|\w|^s}{\w_c^s}
\label{SigmayPert}
\end{eqnarray}

We are now equipped to start understanding the NRG results discussed in Section~\ref{dynamics}.
Owing to the general relation Eq.~(\ref{special}) between spin autocorrelation
functions and Majorana propagators, we restrict the discussion to the latter in
the rest of the paper.
For $\alpha\ll1$ one can safely neglect the small real part of the
self-energy (corresponding to a tiny Knight shift of the spin precession
frequency $\Delta$), so that we get from Dyson's
equations~(\ref{fullx}-\ref{fullz}) for $|\w|<\w_c$:
\begin{eqnarray}
\label{GxG0W0}
\hspace{-0.5cm} \mathcal{I}m
G_x^{G^0W^0}(\w)&\simeq&-\frac{\pi\alpha\w_c^{1-s}}{\w^2}||\w|-\Delta|^s
\theta(\w^2-\Delta^2)\\
\label{GyG0W0}
\hspace{-0.5cm} \mathcal{I}m G_y^{G^0W^0}(\w)&\simeq&-
\frac{\pi\alpha\w_c^{1-s}
|\w|^{2+s}}{(\w^2-\Delta^2)^2+\w^2\left[\pi\alpha\w_c^{1-s}|\w|^s\right]}\\
\label{GzG0W0}
\hspace{-0.5cm} \mathcal{I}m G_z^{G^0W^0}(\w)&\simeq& -
\frac{ \pi\alpha\w_c^{1-s} \Delta^2
|\w|^{s}}{(\w^2-\Delta^2)^2+\w^2\left[\pi\alpha\w_c^{1-s}|\w|^s\right]}
\end{eqnarray}

Following the discussion given in Sec.~\ref{dynamics}, we start by interpreting
the transverse susceptibilities $\mathcal{I}m G_y^{G^0W^0}$ and 
$\mathcal{I}m G_z^{G^0W^0}$.
Clearly the $\delta(\w-\Delta)$ peak is replaced by a sharp resonance, due
to the small but finite lifetime given by $\mathcal{I}m \Sigma_y(\w\simeq\Delta)$
in both expressions~(\ref{GyG0W0}-\ref{GzG0W0}). The low-energy behavior is however
different in those two quantities, as we find $\mathcal{I}m G_y^{G^0W^0}\propto|\w|^{2+s}$ and 
$\mathcal{I}m G_z^{G^0W^0}\propto|\w|^{s}$, in agreement with the NRG results,
see the two topmost panels in Figs.~\ref{NRGChi_s0.1}-\ref{NRGChi_s1.0} and
the related discussion in Sec.~\ref{dynamics}.
A higher density of low-energy modes is indeed obtained in $C_z$ than in $C_y$, due
to the coupling of the bosonic continuum to the spin along the $z$-axis.
This leads therefore to the following decay of the $z$-component of the spin 
autocorrelation function $C_z(t) \propto 1/t^{1+s}$ in the long-time limit.
In the perturbative regime $\alpha\ll1$ and $\alpha\ll\alpha_c$, it is not
surprising to find very accurate agreement between the NRG and our analytical
expressions (note that the real parts of the self-energies have been included
in all plots). Turning to the longitudinal susceptibility $\mathcal{I}m
G_x^{G^0W^0}$, we observe in Eq.~(\ref{GxG0W0}) a broad shoulder for $|\w|>\Delta$, which
is also seen in the numerics, but we obtain instead a {\it hard gap} in the range
$|\w|<\Delta$, in disagreement with the NRG data. This surprising failure of the
perturbative method for $G_x^{G^0W^0}$ at weak dissipation comes from multiparticle 
effects beyond leading order, which do play a qualitative role whenever the spectral density 
is identically zero at lowest order in perturbation theory (otherwise higher order 
self-energy corrections are always small at weak coupling). We now examine this question 
in greater detail.

\subsection{Multiparticle effects}
\label{multiparticle}

\begin{figure}
\includegraphics[scale=0.90]{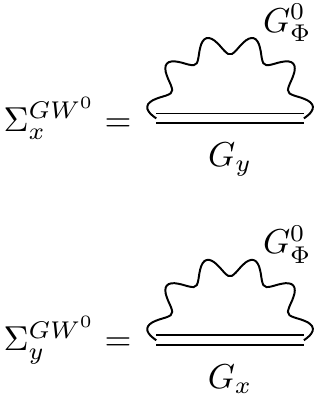}
\vspace{0.1cm}\\
\caption{Selfconsistent fermionic self-energies (analogous to the $GW^0$
approximation), allowing to improve the calculation in frequency ranges 
where lowest order perturbation incorrectly predicts the absence of
spin excitations. This scheme recovers multiparticle excitations that 
fill in the spurious gaps of bare perturbation theory.}
\label{GW0}
\end{figure}

\begin{figure}[ht]
\includegraphics[width=7.0cm]{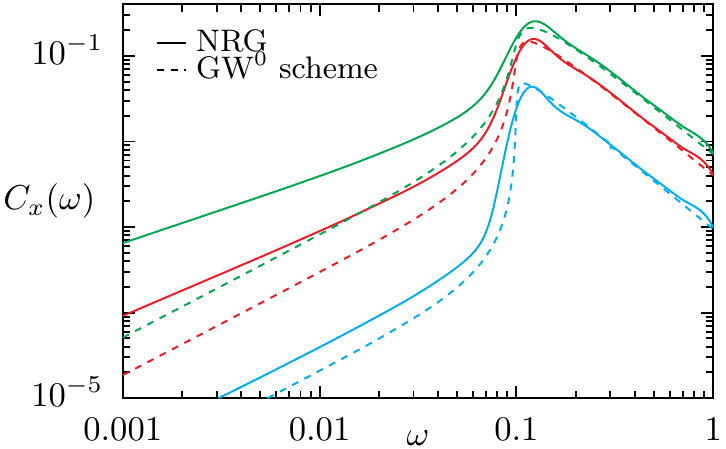}
\caption{Longitudinal spin susceptibility $C_x(\w)$ for $\Delta/\w_c=0.1$, $s=0.1$ and 
$\alpha=0.001,0.004,0.007$ (bottom to top), comparing the NRG result to the 
self-consistent $GW^0$ scheme, allowing to recover the low-energy tail below the 
threshold $\w=\Delta$ (results are quantitatively correct only for the smallest 
$\alpha$ value, see Sec.~\ref{SecVertex} for the improvements that are required near the 
quantum critical point).}
\label{NRGChi_s0.1Iterative}
\end{figure}

We still consider here the spin dynamics at very weak dissipation $\alpha\ll1$ and
$\alpha\ll\alpha_c$, with the goal to improve the diagrammatics for the the lowest order
longitudinal susceptibility $\mathcal{I}m G_x^{G^0W^0}$, whose spectrum is
spuriously gapped at lowest order in perturbation theory.
This inconsistency can be easily tracked to expression~(\ref{barex}), where the bare 
{\it transverse} propagator $G_y^0$ resumes to a $\delta(|\w|-\Delta)$ peak and misses 
the $\w^{2+s}$ low-energy modes obtained from the coupling to the bosonic bath, 
see equation~(\ref{GyG0W0}). The discrepancy is thus resolved by reinjecting the 
full Majorana-Green's functions $G_i$ into the Majorana self-energies (this is the 
so-called $GW^0$ scheme, see Fig.~\ref{GW0}):
\begin{eqnarray}
\label{SxGW0}
\mathcal{I}m \Sigma_x^{GW^0}(\w) & = & \alpha \int_0^\w
\frac{\mathrm{d}\epsilon}{\pi} \mathcal{I}m G_y(\epsilon) \mathcal{I}m
G_\Phi^0(\w-\epsilon) \\
\label{SyGW0}
\mathcal{I}m \Sigma_y^{GW^0}(\w) & = &\alpha \int_0^\w
\frac{\mathrm{d}\epsilon}{\pi} \mathcal{I}m G_x(\epsilon) \mathcal{I}m
G_\Phi^0(\w-\epsilon)
\end{eqnarray}
The low-energy tails of the transverse susceptibility will therefore fill in
the gap of the longitudinal susceptibility, so that in principle a single
iteration of the above self-consistent equation should be sufficient required. 
The results are shown in Fig.~\ref{NRGChi_s0.1Iterative}, demonstrating our correct
physical interpretation of the missing low energy tails in $G_x$. We note that 
the two transverse susceptibilities are very mildly affected by the self-consistency 
at weak coupling, and were therefore not shown.

A second instance where multiparticle effects are relevant occurs in the
frequency range above the high energy cutoff, now for all three spin
correlations.
Indeed, lowest order perturbation theory provides again a gap for excitations
with $|\w|>\w_c$, see Eqs.~(\ref{SigmaxPert}-\ref{SigmayPert}) in the case of
one of the transverse spin correlation function (similar results are obtained
for the other spin susceptibilities).
\begin{figure}[ht]
\includegraphics[width=7.0cm]{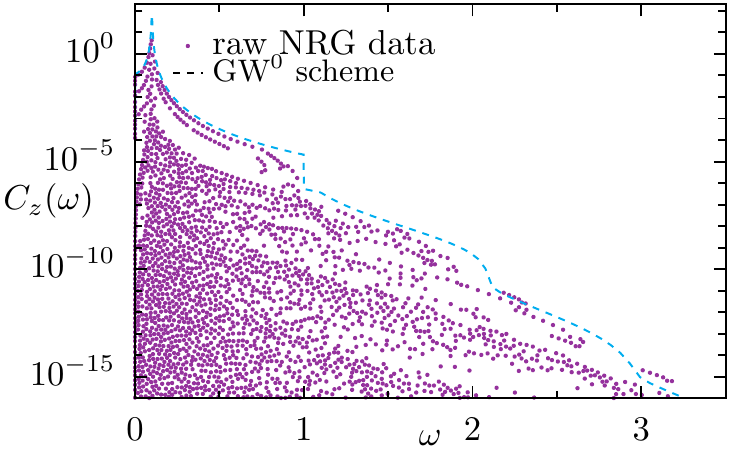}
\caption{Transverse spin susceptibility $C_z(\w)$ for $\Delta/\w_c=0.1$, $s=0.1$ and 
$\alpha=0.002$, comparing the raw NRG data to the self-consistent $GW^0$ scheme.
In contrast to bare perturbation theory, which leads to a gap for $\w>\w_c$, the
self-consistent resummation correcly describes the tail arising from multiparticle excitations 
above the threshold $\w_c$ (note the linear frequency scale).}
\label{NRGChi_s0.1Above}
\end{figure}
In contrast, the NRG data show a tail with small yet non-zero spectral weight above 
the cutoff $\w_c$, see Fig.~\ref{NRGChi_s0.1Above}. This continuum of magnetic excitations
decays very quickly at increasing energy, and displays furthermore well-defined
thresholds, pointing again to multiparticle effects, where more and more bosons
are exchanged with the spin degrees of freedom, in agreement with recent
observation.~\cite{Nalbach} Using the fermionic
self-consistent scheme proposed in Eqs.~(\ref{SxGW0}-\ref{SyGW0}), we indeed
reproduce numerically both the rapidly decaying tail and the multiparticle
thresholds at quantized values of the cutoff $\w_c$ (we note that the
multiparticle thresholds seen in the raw NRG data shift do not precisely match
the quantized values, possibly an artifact of the discretization procedure). 
This final comparison concludes our study of the weak coupling regime 
$\alpha\ll\alpha_c$ of the spin-boson model.

\section{Dynamics near the quantum critical point}
\label{QCP}

\subsection{Role of the bosonic mass}

We provide here general considerations on the structure of perturbation
theory in terms of the coupled field theory of fermions and bosons,
with the emphasis on the regime near the quantum critical point.
The main motivation is to understand precisely where the singularities
associated to critical modes will occur in the diagrammatics. 
From the knowledge gained in the previous
weak coupling analysis, we understand that the fermionic sector is
immune to singularities as long as the bosonic propagator remains
regular. Clearly self-consistency for the fermionic self-energies ($GW^0$ scheme) 
provides only small perturbative correction and cannot account for critical
behavior. One can convince oneself that the frequency-dependent vertex $\Gamma$ 
(see Fig.~\ref{SigmaExact}) is also regular as dissipation is increased.
The only possibility left is that only the bosonic propagator encapsulates
the singular dynamics via its self-energy (see general diagrammatic
expression on the lower panel in Fig.~\ref{SigmaExact}).
Indeed using the exact Dyson's equation~(\ref{fullPhi})
and the asymptotic bare propagator~(\ref{ImGPhi}-\ref{ReGPhi}), we
get the low frequency retarded bosonic Green's function:
\begin{eqnarray}
\nonumber
G_\Phi(\w) &=& \frac{1}{m_\Phi+[a+ib\mathrm{Sign}(\w)]|\w|^s-\Sigma_\Phi(\w)+\Sigma_\Phi(0)}\\
\label{critical}\\
\label{mass}
m_\Phi &=& \frac{s}{4\w_c}-\Sigma_\Phi(0)
\end{eqnarray}
with $m_\Phi$ the bosonic mass, $a$ and $b$ real coefficients that can be read
from Eqs.~(\ref{ImGPhi}-\ref{ReGPhi}).
This simple expression allows to understand the development of critical 
fluctuations in the spin-boson model: the quantum critical point is just
associated to a vanishing of the mass $m_\Phi$ in the denominator of the 
bosonic propagator, due to the progressive building up of the bosonic self-energy. 
At the quantum critical point,
$\Sigma_\Phi(0)=s/(4\w_c)$, and the correlation function now diverges
as $G_\Phi(\w)\propto |\w|^{-s}$ at low frequency, owing to expression~(\ref{critical}).
One can check that the frequency dependence of the bosonic self-energy
$\Sigma_\Phi(\w)$ is subdominant with respect to $\w^s$ in all orders in 
perturbation theory, so that the $-s$ exponent of the critical fluctuations turns 
out to be {\it exact}.
In contrast, the bosonic Green's function has a non-divergent power law behavior 
at low frequency $G_\Phi(\w)-G_\Phi(0)\propto |\w|^{s}$ in the whole delocalized phase. 
This change of exponent near the quantum critical point can be used to explain
the behavior of the transverse spin susceptibility, calculated previously with the
NRG. 
\begin{figure}
\includegraphics[scale=0.90]{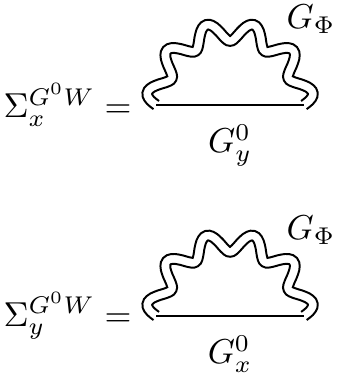}
\vspace{0.1cm}\\
\caption{Fermionic self-energies (without self-consistency) including a dressed
bosonic propagator (this is analogous to the $G^0W$ approximation). This scheme
is necessary to recover the critical modes near the quantum critical point, but
asks for a careful evaluation of the bosonic self-energy (see
Fig.~\ref{TwoLoops} and the related discussion in Sec.~\ref{SelfTwoLoops} and 
Sec.~\ref{SelfLadder}).}
\label{G0W}
\end{figure}
When using the full bosonic propagator within the fermionic self-energy 
(see Fig.~\ref{G0W}), we get:
\begin{eqnarray}
\mathcal{I}m \Sigma_y^{G^0W}(\w) & = & -\frac{1}{2} \mathrm{Sign}(\w) 
\mathcal{I}m G_\Phi(\w)\\
& \propto &  |\w|^{-s} \;\;\mathrm{at}\;\;\alpha=\alpha_c
\end{eqnarray}
From the exact Dyson equation for the $\eta_z$ Majorana Green's function 
Eq.~(\ref{fullz}), we then get at low frequency:
\begin{eqnarray}
\mathcal{I}m G_z^{G^0W}(\w) & \simeq & \frac{1}{\Delta^2} 
\mathcal{I}m \Sigma_y^{G^0W}(\w) \\
& \propto & |\w|^{-s} \;\;\mathrm{at}\;\;\alpha=\alpha_c
\end{eqnarray}
The divergence of the transverse spin susceptibility at the quantum critical
point is thus explained (again, we stress that the above exponent is exact).
Considering also the second transverse susceptibility Eq.~(\ref{fully}), we get
at low frequency:
\begin{eqnarray}
\mathcal{I}m G_y^{G^0W}(\w) & \simeq & \frac{\w^2}{\Delta^4} 
\mathcal{I}m \Sigma_y^{G^0W}(\w) \\
& \propto & |\w|^{2-s} \;\;\mathrm{at}\;\;\alpha=\alpha_c
\end{eqnarray}
as also observed with the NRG calculations of Sec.~\ref{dynamics}.

This discussion shows that even in the weak dissipation regime $\alpha\ll1$, the
presence of such critical modes for $\alpha\lesssim\alpha_c$ will invalidate
bare perturbation theory. It is thus necessary to investigate in more detail
the bosonic correlations, in order to improve the agreement between numerics
and diagrammatic calculation. We start by an analytic computation of the bosonic
self-energy up to two-loop order.

\subsection{Critical line at two-loop order}
\label{SelfTwoLoops}

\begin{figure}
\includegraphics[scale=0.70]{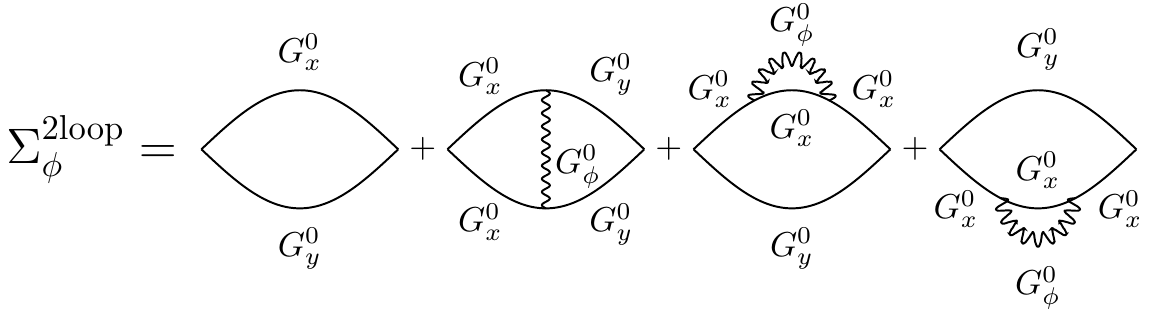}
\vspace{0.1cm}\\
\caption{Diagrammatic representation for the bosonic self-energy computed up 
to two-loop order, leading to Eq.~(\ref{TwoLoopFinal}).}
\label{TwoLoops}
\end{figure}

The lowest order (one-loop) contribution to the bosonic self-energy (first
diagram in Fig.~\ref{TwoLoops}) is readily computed:
\begin{equation}
\label{SPhi1loop}
\Sigma_\phi^\mathrm{1loop}(\w) = \frac{\alpha}{2\Delta}
\frac{\Delta^2}{\Delta^2-(\w+i0^+)^2}.
\end{equation}
This level of approximation in the diagrammatics is equivalent to the
standard RPA for the interacting electron-gas.~\cite{Vignale}
From vanishing of the mass in Eq.~(\ref{mass}), the above result gives 
the critical dissipation at one-loop order
\begin{equation}
\alpha_c^\mathrm{1loop} = \frac{s}{2} \frac{\Delta}{\w_c}+\mathcal{O}(s^2),
\label{alphaOneLoop}
\end{equation}
recovering previous results.~\cite{Vojta} We recognize here the multi-scale
nature of the perturbative expansion in the dissipation strength. While
the high-energy part of the spectrum is well described as long as $\alpha\ll1$,
the low energy sector will be immune to quantum critical fluctuations as long
as $\alpha\ll\alpha_c$.
While $\alpha_c$ vanishes for $s\to0$, showing that the quantum critical
point is perturbatively accessible, the above result is only accurate
for very small $s$. As we will demonstrate now, an anomalous power-law also
shows up in the dependence of $\alpha_c$ with respect to the 
dimensionless parameter $\Delta/\w_c$. To establish this result, we
need to push the weak-coupling calculation of the bosonic mass
to next to leading order (two-loops), see Fig.~\ref{TwoLoops}.
After straigthforward but lengthy calculations, we find at zero
frequency:
\begin{eqnarray}
\nonumber
\Sigma_\phi^\mathrm{2loop}(0) &=& \frac{\alpha}{2\Delta}
+\frac{\alpha^2}{4\Delta^2} \int_0^{+\infty} 
\!\! \frac{d\epsilon}{\pi} \, \mathcal{I}m G_\Phi^0(\epsilon) \,
\frac{3\Delta+2\epsilon}{(\Delta+\epsilon)^2}\\
&\simeq& \frac{\alpha}{2\Delta}+
\frac{\alpha^2}{s}\frac{\w_c}{\Delta^2}\left[1-\left(1-\frac{s}{2}\right)\left(\frac{\Delta}{\w_c}\right)^s\right]
\label{TwoLoopFinal}
\end{eqnarray}
where the last equation applies for $\Delta\lesssim\w_c$.
We now solve the equation 
$s/(4\w_c)=\Sigma_\phi^\mathrm{1loop}(0)+\Sigma_\phi^\mathrm{2loop}(0)$
determining the quantum critical point, which gives in the small $s$ limit:
\begin{equation}
\label{selftwoloop}
\alpha_c^\mathrm{2loop}(\w) =
\frac{s}{2}\frac{\Delta}{\w_c}
\left[1+\frac{s}{2}-s\log\left(\frac{\Delta}{\w_c}\right)\right]
+\mathcal{O}(s^3).
\end{equation}
This precise form confirms that the weak-coupling diagrammatics is controlled 
as a systematic small-$s$ expansion.
The presence of a logarithmic correction hints for the following anomalous
power-law behavior:
\begin{equation}
\alpha_c^\mathrm{2loop}(\w) \simeq
\frac{s}{2} \left(1+\frac{s}{2}\right)
\left(\frac{\Delta}{\w_c}\right)^{1-s}.
\label{alphac}
\end{equation}
We will first examine this analytical result with respect to the NRG
phase diagram, before turning to a more rigorous derivation.

\subsection{Phase diagram of the sub-ohmic spin-boson model}

\begin{figure}
\includegraphics[scale=1.0]{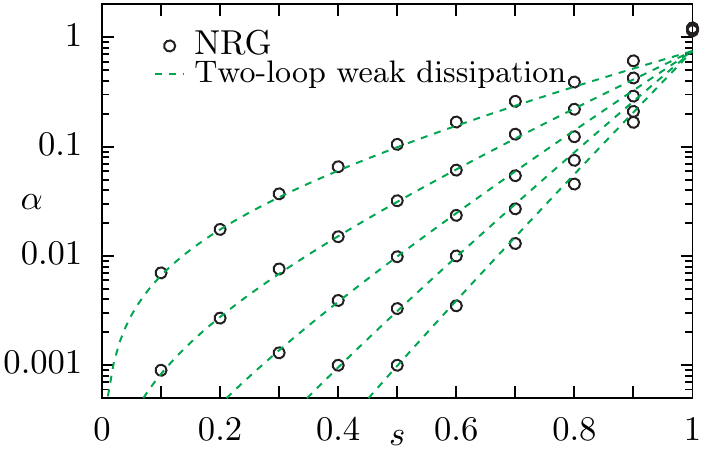}
\vspace{0.1cm}\\
\caption{Phase diagram in the ($s$,$\alpha$) plane for 
$h\equiv\Delta/\w_c=0.00001,0.0001,0.001,0.01,0.1$ (bottom to top), comparing
the NRG simulations versus the two-loop weak dissipation result (valid for $s\ll1$).}
\label{CompareAll}
\end{figure}

According to the above result, the formula~(\ref{alphac}) for the critical 
dissipation $\alpha_c$ remains perturbative either for small $s$
(at arbitrary $\Delta/\w_c\lesssim1$) or for small $\Delta/\w_c$.
In the latter limit, $s$ can take relatively large values, but should 
not be too close to 1, otherwise the small coefficient $(\Delta/\w_c)^{1-s}$
becomes of order 1, and perturbation theory breaks down. In this case,
an alternative approach, based on a development at small $h\equiv\Delta/\w_c$ 
but finite $\alpha$ can be put forward.~\cite{Kosterlitz,Bulla} This results
in the well-known one-loop scaling equations (which were initially obtained from the
long-range unidimensional Ising model):
\begin{eqnarray}
\label{dalpha}
\frac{d\alpha}{d l} &=& (1-s)\alpha-\alpha h^2 \\
\label{dh}
\frac{dh}{d l} &=& (1-\alpha)h 
\end{eqnarray}
where the parameter $l$ characterizes the change of scale during the
renormalization procedure.
We stress that these flow equations are valid for all values of $\alpha$, as 
long as $h\ll1$, and are thus complementary to the regime of validity
of our small $\alpha$ expansion at fixed $h$.
Combining both equations, one obtains:
\begin{equation} 
\frac{d}{d l}\left[\log(\alpha)-\alpha+\frac{h^2}{2}-(1-s)\log(h)\right]=0
\end{equation}
From Eqs.~(\ref{dalpha}) and~(\ref{dh}) the renormalization fixed point occurs at
$\alpha^\ast=1$ and $h^\ast=\sqrt{1-s}$ for $s\leq1$, so that the critical line
$\alpha_c(h)$ is obtained by the condition:
\begin{equation} 
\log(\alpha_c)+(1-\alpha_c)+\frac{h^2-(1-s)}{2}-(1-s)\log\left(\frac{h}{\sqrt{1-s}}\right)=0
\label{KTflow}
\end{equation}
leading to the behavior $\alpha_c\propto h^{1-s}$ at $h\ll1$ for all $s$. We also recover 
from this analysis the exact value $\alpha_c=1$ for $s=1$ in the $h\to0$ limit.
\begin{figure}
\includegraphics[scale=1.0]{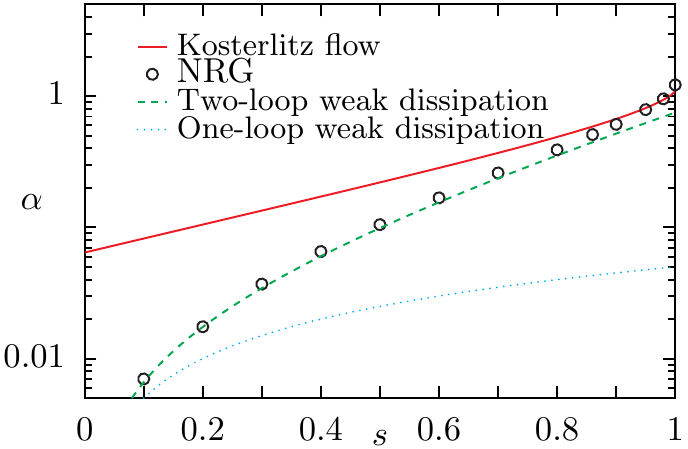}
\vspace{0.1cm}\\
\caption{Phase diagram in the ($s$,$\alpha$) plane for $h\equiv\Delta/\w_c=0.1$,
comparing the NRG data to the Kosterlitz flow obtained from the resolution of
Eq.~(\ref{KTflow}) (giving reliable results for $0.9<s<1$), and to the one-loop
and two-loop formulas~(\ref{alphaOneLoop}) and (\ref{alphac}) (the latter is accurate 
on the wide range $0<s<0.8$).}
\label{CompareDelta0.1}
\end{figure}
A joint comparison of the small $\alpha$ expansion and of the Kosterlitz flow to
the NRG data is provided in Fig.~\ref{CompareDelta0.1} for the case $h=0.1$. 
While the two expansions work where they are expected to, the two-loop small 
$\alpha$-development is quantitatively accurate on the wide range $0<s<0.8$, 
while the one-loop small $h$-development is restricted to the small range $0.9<s<1$. 
Although pushing the small $h$ expansion to higher order would enhance the range
of validity of the method, it is unlikely that this would recover the expected
vanishing of $\alpha_c\simeq (s/2)h$ at small $s$. 
In contrast, the small $\alpha$-expansion converges very rapidly even
for the largest $s$ values: the extreme case $s=1$ leads from Eq.~(\ref{alphac}) to 
the successive estimates $\alpha_c^\mathrm{1loop}=1/2$ and
$\alpha_c^\mathrm{2loop}=3/4$ in the $\Delta/\w_c\ll1$ limit, approaching the
exact value $\alpha_c=1$ in a steady fashion. As a clear illustration of this
statement, we finally present in Fig.~\ref{CompareAll} a more systematic study of 
the phase diagram for several $h$ values, showing that the small $\alpha$
expansion remains accurate even at vanishing $h$, provided that $s<0.8$.

\subsection{Ladder resummation in the bosonic self-energy}
\label{SelfLadder}

\begin{figure}
\includegraphics[scale=0.67]{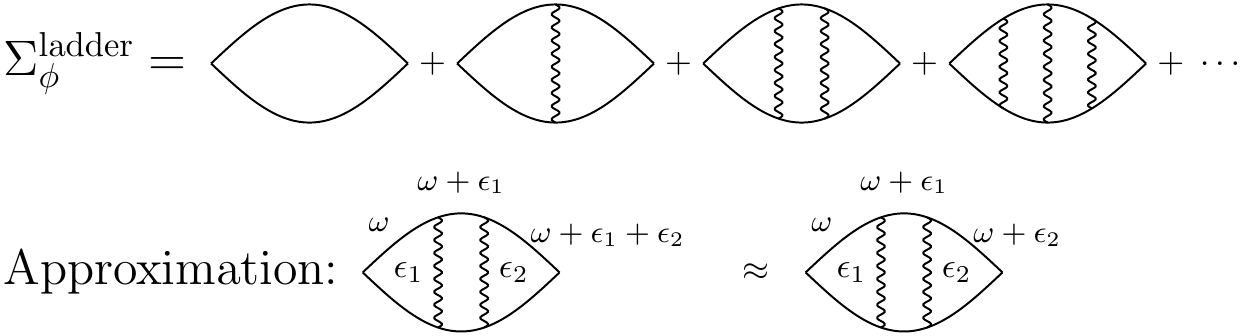}
\vspace{0.1cm}\\
\caption{Upper panel: ladder series for the bosonic self-energy that is needed
to obtain the correct phase boundary within the diagrammatic calculation. Lower
panel: single frequency approximation used at the two-rung level (and similarly for
higher order contributions), see text.}
\label{ladder}
\end{figure}

We have just checked that our analytical expression for the critical 
dissipation~(\ref{alphac}) is in excellent agreement with the NRG data.
However, the derivation of this result was performed by a {\it strict} expansion at
two-loops, and a re-exponentiation of the logarithmically singular terms.
In the spirit of a numerical scheme based only on Green's functions, such
as developed in Sec.~\ref{Majorana}, the correct phase boundary will not
be recovered if strict fourth order bosonic self-energies are used, and
a clever resummation scheme should be found, that would be a diagrammatic
equivalent to the re-exponentiation of the analytical result. Insight on
this issue can be gained from the analytical derivation, because the 
logarithmic term in Eq.~(\ref{selftwoloop}) can be traced back to a single
diagram at two-loop order, namely the one with a single-rung ladder exchange in Fig.~\ref{TwoLoops}
(the last two diagrams with Majorana self-energy corrections are regular, and
provide the perturbative corrections of order $s^2$ in Eq.~(\ref{alphac})). This hints that the
correct bosonic self-energy, and hence a quantitative phase diagram, will be
obtained upon performing a ladder resummation to all orders, similar to
the diffuson mode in disordered systems,~\cite{DiCastro} see Fig.~\ref{ladder}.
Performing such calculation, even numerically, turns out to be difficult,
because this requires the self-consistent determination of a four-electron vertex 
(depending of three independent external frequencies). A standard approximation 
in the field of disordered systems,~\cite{DiCastro} but also in the context of 
interacting electron liquids~\cite{Vignale} (called the Ng-Singwi scheme, or $GW_\mathrm{eff}$ 
approximation) amounts to keeping a single running frequency within the vertex, see
lower panel of Fig.~\ref{ladder} in the case of a two-rung ladder. In our case,
this approximation is vindicated by the fact that the low frequency part of
the bosonic Green's function provides the most important contribution to the
Feynman diagram, so that strict energy conservation can be omitted from one rung to the
next.
We stress again that the inclusion of such vertex corrections in the fermionic
self-energies are regular, and can be neglected at weak coupling, contrary
to the singular bosonic propagator that we treat here.
The ladder series including the full three particle vertex can now be explicitely carried
out for the zero-temperature Matsubara bosonic self-energy:
\begin{equation}
\Sigma_\Phi^\mathrm{ladder}(i\nu) \simeq \alpha \int \frac{d\w}{2\pi} G_y^0(i\w+i\nu)
G_x^0(-i\w) \frac{1}{1-\Lambda^0(i\w,i\nu)}
\end{equation}
where 
\begin{equation}
\Lambda^0(i\w,i\nu) \simeq \alpha \int \frac{d\epsilon}{2\pi} G_\Phi^0(i\epsilon)
G_y^0(-i\w-i\epsilon)
G_x^0(i\w+i\nu+i\epsilon).
\end{equation}
Evaluation of the zero-frequency self-energy can be carried out analytically 
in the small $\Delta/\w_c$ limit, leading to:
\begin{equation}
\Sigma_\Phi^\mathrm{ladder}(0) = \frac{\alpha}{2\Delta}
\frac{1}{1+\frac{2\alpha\w_c}{s\Delta}\left[1-\left(\frac{\Delta}{\w_c}\right)^s\right]}.
\label{SPhiFinal}
\end{equation}
The phase boundary can now be obtained by solving
$\Sigma_\Phi^\mathrm{ladder}(0)=1/G_\Phi^0(0)$, leading to the critical value
$\alpha_c^\mathrm{ladder}=(s/2)(\Delta/\w_c)^{1-s}$, in agreement with
Eq.~(\ref{alphac}). Note that the small corrections at order $s^2$ are missing in the 
above derivation, because fermionic self-energy insertions were discarded in the
present calculation (these were computed explicitely in Sec.~\ref{SelfTwoLoops}).
This correct result for the critical dissipation confirms that the ladder resummation 
is the good strategy to reproduce the correct phase boundary in a fully diagrammatic 
calculation. We finish the paper by completing this study of the phase diagram, which will 
require a careful discussion of vertex corrections in the fermionic self-energies as well.

\subsection{Full Majorana diagrammatics and the three-particle vertex}
\label{SecVertex}

\begin{figure}[ht]
\includegraphics[width=7.0cm]{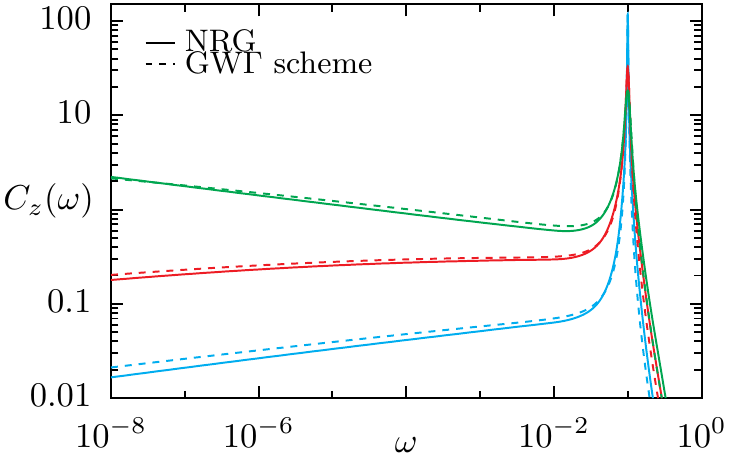}
\includegraphics[width=7.0cm]{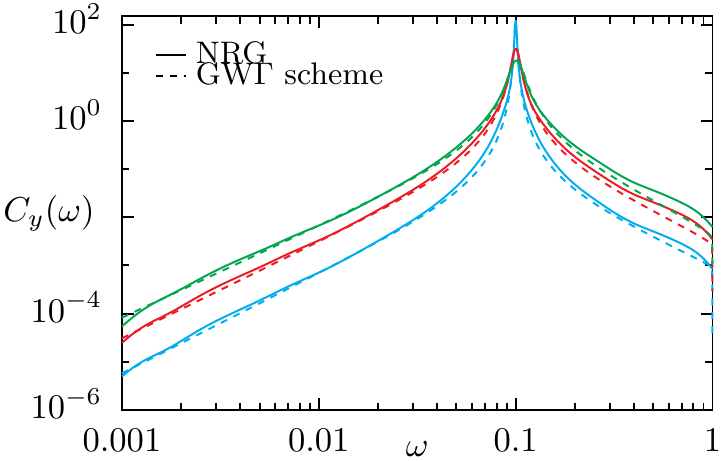}
\includegraphics[width=7.0cm]{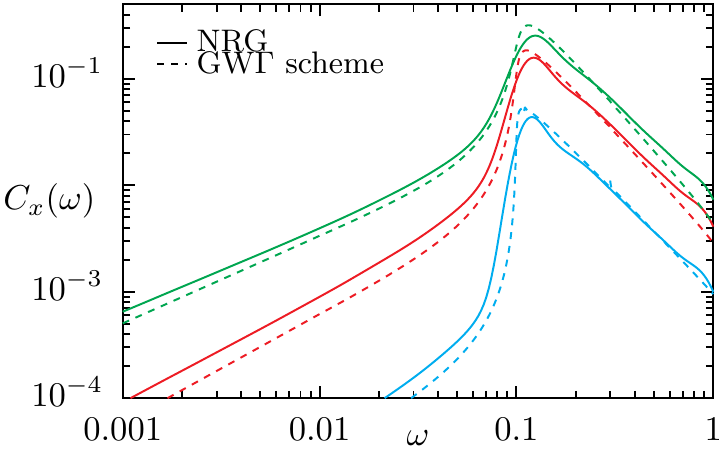}
\caption{Spin susceptibilities $C_z(\w)$, $C_y(\w)$ and $C_x(\w)$
(top to bottom panels) for $\Delta/\w_c=0.1$, $s=0.1$ and increasing
dissipation strength $\alpha=0.001,0.004,0.007$ (bottom to top), compared to the full
Majorana diagrammatics (GW$\Gamma$ scheme).} 
\label{NRGFinal_s0.1}
\end{figure}

\begin{figure}[ht]
\includegraphics[width=7.0cm]{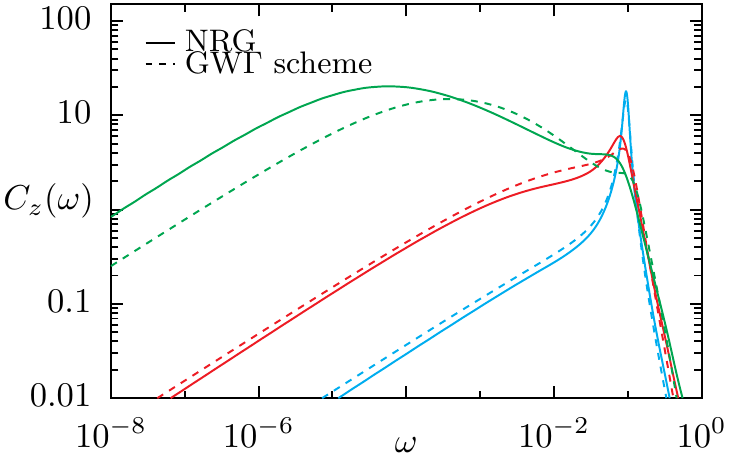}
\includegraphics[width=7.0cm]{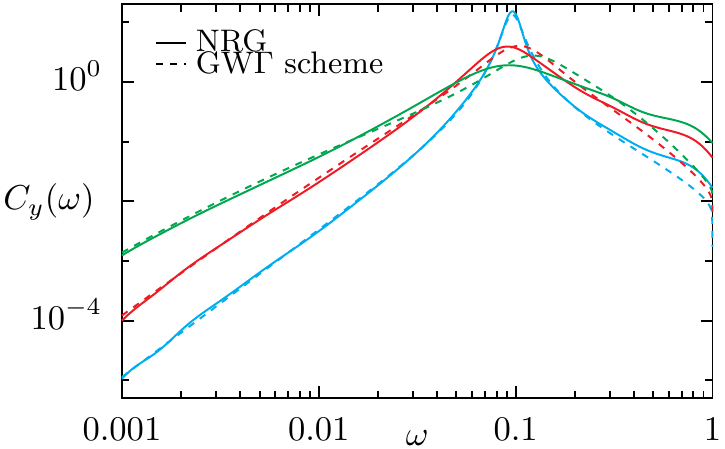}
\includegraphics[width=7.0cm]{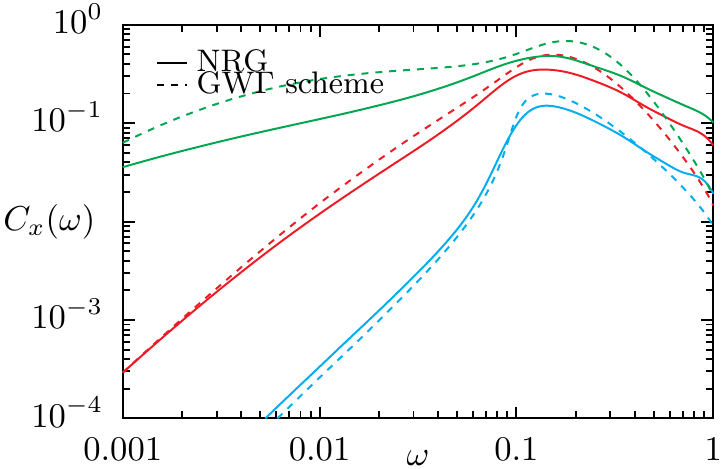}
\caption{Spin susceptibilities $C_z(\w)$, $C_y(\w)$ and $C_x(\w)$
(top to bottom panels) for $\Delta/\w_c=0.1$, $s=0.5$ and increasing
dissipation strength $\alpha=0.02,0.06,0.1$ (bottom to top), compared 
to the full Majorana diagrammatics (GW$\Gamma$ scheme).} 
\label{NRGFinal_s0.5}
\end{figure}

We are now equipped to collect all the needed ingredients for a
successful implementation of the Majorana diagrammatics up to the
quantum critical point (as long as $\alpha\ll1$, i.e. for values of
$s$ not too close to the ohmic regime). The next step beyond the previous
levels of approximation (G$^0$W$^0$ and GW$^0$ schemes) is the
inclusion of the vertex corrections (at ladder level) in the bosonic 
self-energy Eq.~(\ref{SPhiFinal}), which is required to obtain the correct
phase boundary and to describe quantitatively the critical low-energy fluctuations.
We note that the GW scheme with fully self-consistent propagators but without
vertex corrections does not recover the correct critical point, and will not
be further considered. 
We therefore propose a GW$\Gamma$ approximation, which should incorporate the vertex 
corrections {\it both} for the bosonic and fermionic 
self-energies (see Fig.~\ref{SigmaExact}). 
Indeed, in order to emphasize the importance of the vertex corrections in the
fermionic sector, let us first rewrite the bosonic self-energy as 
$\Sigma_\Phi^\mathrm{ladder}(\w)\simeq\alpha^R/(2\Delta)$, introducing a renormalized 
coupling
\begin{equation}
\alpha^R=\sqrt{\alpha} \Gamma\simeq
\frac{\alpha}{1+\frac{2\alpha\w_c}{s\Delta}\left[1-\left(\frac{\Delta}{\w_c}\right)^s\right]}
\label{alphaR}
\end{equation}
in terms of the static three-particle vertex $\Gamma$ (see Fig.~\ref{SigmaExact})
computed at the ladder level.
\begin{figure}[ht]
\includegraphics[width=7.0cm]{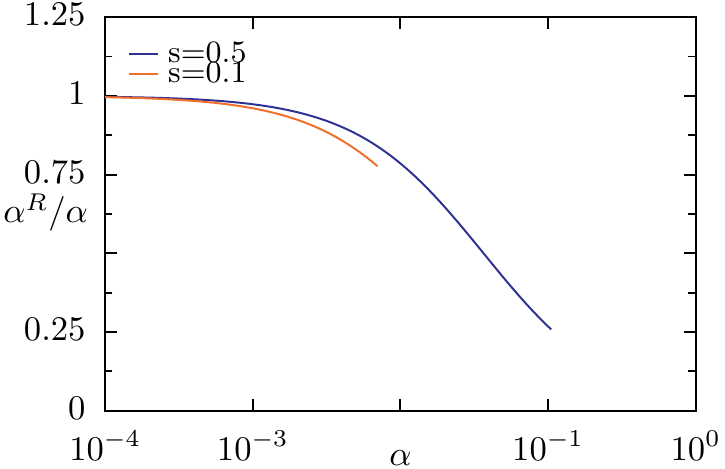}
\caption{Renormalization factor $\alpha^R/\alpha$ of the dissipation strength 
from Eq.~(\ref{alphaR}) computed for $\Delta/\w_c=0.1$ as a function of $\alpha$ up to 
the critical point ($\alpha_c=0.007,0.105$ respectively for the two values 
$s=0.1,0.5$).}
\label{FigalphaR}
\end{figure}
This quantity is plotted in Fig.~\ref{FigalphaR} for two values of $s$ as
a function of $\alpha<\alpha_c$, which shows that the full vertex $\Gamma$ remains
finite up to the critical point, in agreement with our general expectation that
this quantity remains regular at all orders in perturbation theory.
Interestingly, the renormalization of the coupling constant $\alpha$ can be
relatively large, up to 20\% and 75\% for the two cases $s=0.1$ and $s=0.5$
considered here. Such a sizeable effect will impact as well the fermionic
self-energies, so that our consistent GW$\Gamma$ scheme will finally rely on the
following set of equations:
\begin{eqnarray}
\label{SxGWGamma}
\hspace{-.5cm}
\mathcal{I}m \Sigma_x^{GW\Gamma}(\w) & = & \alpha^R \int_0^\w
\frac{\mathrm{d}\epsilon}{\pi} \mathcal{I}m G_y(\epsilon) \mathcal{I}m
G_\Phi(\w-\epsilon) \\
\label{SyGWGamma}
\hspace{-.5cm}
\mathcal{I}m \Sigma_y^{GW\Gamma}(\w) & = &\alpha^R \int_0^\w
\frac{\mathrm{d}\epsilon}{\pi} \mathcal{I}m G_x(\epsilon) \mathcal{I}m
G_\Phi(\w-\epsilon)\\
\label{SPhiGWGamma}
\hspace{-.5cm}
\Sigma_\Phi^{GW\Gamma}(\w) & = & \frac{\alpha^R}{2\Delta}
\end{eqnarray}
with $\alpha^R$ given in Eq.~(\ref{alphaR}).
The numerical implementation of the above equation, with self-consistency
imposed by Dyson's equations~(\ref{fullx}-\ref{fullPhi}),
is readily implemented, and gives the results displayed in
Figs.~\ref{NRGFinal_s0.1} and \ref{NRGFinal_s0.5} (we note again that
fermionic self-consistency is only used in order to fill in the
spurious gap of $C_x$, but it is not too important here).
%
In the case $s=0.1$, the outcome of the Majorana diagrammatics is at par 
with the NRG results for all three spin susceptibilities, both at low 
and high energies, and including the quantum critical regime. 
For the larger $s=0.5$ value, agreement is still quite good, and the 
visible deviations in the quantum critical regime (for $\alpha=0.1$) arise 
from slight discrepancies in the critical $\alpha_c$ between the diagrammatics 
and the NRG, so that the crossover scale is not accurately captured.
We emphasize that this quantitative description of the spin dynamics
relies of the weak coupling nature of the spin boson model for small $s$
values. The key step here was to understand that the most important
contributions to the diagrammatics are the renormalization (beyond leading
order) of the bosonic mass shift and the vertex corrections. It is possible that
analytical renormalization group techniques (such as developed in
Ref.~\onlinecite{Fritz} for the quantum phase transition in the pseudogap Kondo 
problem at weak coupling) may be applied to the spin-boson Hamiltonian based
on the present understanding, but this would likely require more complex machinery to
obtain similar results.
This successful comparaison seals our study of the spin dynamics in delocalized
phase of the spin-boson model.

\section{Conclusion}

We have investigated the {\it equilibrium} spin dynamics of the sub-ohmic spin-boson model, which
generically describes a two-level quantum system coupled to a dissipative environment of harmonic 
oscillators (characterized by a power-law spectrum $J(\w)=2 \pi \alpha \w_c^{1-s} 
\w^s\theta(\w)\theta(\w_c-\w)$ with $s\leq 1$) and submitted to a perpendicular magnetic field 
$\Delta$.
Two complementary techniques were employed, that were quantitatively benchmarked
against each other. First, the bosonic Numerical Renormalization Group (NRG) was used 
with an optimized broadening method (``b-trick'') to obtain reliably the dynamical spin 
susceptibilities from weak ($\alpha\ll1$) to strong ($\alpha\simeq 1$) dissipation.
Here, we considered not only the usual transverse spin susceptibility (orthogonal
to the field and parallel to the bosonic bath), but also for the first time the longitudinal 
(parallel to the field) and the second transverse (orthogonal to both bath and field) ones.
Second, we established a general Majorana-fermion diagrammatic perturbation theory, controlled in
the weak dissipation limit $\alpha\ll1$. Because the spin-localization quantum phase transition 
takes place at a critical value $\alpha_c$ which becomes perturbatively
accessible ($\alpha_c\ll1$) either in the small field limit ($\Delta\ll\w_c$ for
all $s<1$) or in the strong sub-ohmic regime ($s\ll1$ for all $\Delta/\w_c$), it
was possible to extend the diagrammatic method up to the quantum critical point.
We then examined precisely how various flavors of GW-like approximations perform against the
numerically exact results, in two different regimes.
For $\alpha\ll\alpha_c$, bare perturbation theory (analogous to the standard Bloch-Redfield
approximation for the non-equilibrium dynamics) was sufficient to describe well the finite energy dissipative 
features near the applied magnetic field $\Delta$, but fermionic
self-consistency was needed in order to improve both on the low and high energy parts 
of the spectrum. Self-consistency was required because some of the bare self-energies 
can show spurious gaps that are in reality filled by multiparticle excitations.
For $\alpha\lesssim\alpha_c$, huge deviations appeared in the low-energy limit between 
lowest order perturbation theory and the NRG data, due to the proximity to
the quantum critical point. We showed that the effective interaction mediated by
the bosonic field was the crucial ingredient to correctly describe the physics
close to criticality, and that it should include vertex corrections (at the
ladder level) in the bosonic and fermionic self-energies for quantitative agreement 
with the numerics. 
Such diagrammatic contributions are difficult to evaluate for higher dimensional field 
theories of coupled fermions and bosons~\cite{Hertz,Vignale} and are rarely considered 
(see however Ref.~\onlinecite{Rech}). 
A two-loop analysis allowed us also to obtain a simple equation for the critical line,
that describes the NRG phase diagram very accurately in a wide range of parameters.
We also presented a proof of Shiba's relation generalized to the sub-ohmic case
and the derivation of an effective bosonic action that may support the validity
of the quantum-to-classical mapping.

We finally discuss possible outlooks for the methodology and the ideas developed here. 
First, the addition of a magnetic field component $\epsilon S^z$ parallel to the quantization 
axis may be considered with special interest~\cite{LeHur}. Indeed, understanding the magnetization
process is the next step towards a good description of the localized phase of
the spin-boson model, which was non considered in the present work, and which has led to 
recent debate.~\cite{Vojta,Kirchner,Rieger,Erratum,Anders,Feshke,Zhang,TongTrunc}
Second, further extensions of the model to multiple bosonic environments present very
interesting physics in the ohmic case,~\cite{Novais} which remains to be investigated 
in greater generality. Third, an important consequence of our results is that
the long time equilibrium dynamics is never accurately captured by lowest order perturbation
theory in the coupling between the spin and the environment (unless dissipation is
vanishingly small). This remark should have deep implications for the out-of-equilibrium 
long-time dynamics in dissipative models, which ought to be re-examined beyond
the Bloch-Redfield regime. Extension of the Majorana diagrammatic method to two-times Keldysh
Green's function could be also considered in the context of sudden quenches in
the spin boson model near the quantum critical point.
Lastly, considering the spin-boson Hamiltonian as a toy model for quantum criticality, 
the quantitative understanding achieved here may be relevant for field theories of electrons 
coupled to bosons in higher dimensions.

\acknowledgments{We wish to thank R.~Bulla, F.~Evers, L.~Fritz, A.~Shnirman, N.-H.~Tong, T.~Vojta, 
and R.~Whitney for useful discussions, and especially M.~Vojta for many stimulating exchanges. 
We acknowledge financial support from the ERC Advanced Grant MolNanoSpin n°226558, from MIUR under 
the FIRB IDEAS Project No. RBID08B3FM, and by the “Universita Italo Francese/Universit\'e Franco 
Italienne” (UIF/UFI) under the Program VINCI 2008 (Chapter II).}

\appendix

\section{Shiba's relations for the sub-ohmic spin-boson model}
\label{shiba}
The goal of this appendix is to provide a simple derivation of
Shiba's relation Eq.~(\ref{shibasubohmic}) generalized to the sub-ohmic
case. The first step of the proof is to note that the linear coupling 
between the $z$ component of the spin and the bosonic continuum, see
Hamiltonian Eq.~(\ref{ham}), implies an exact relation between the longitudinal 
spin susceptibility and the full retarded bosonic Green's function:
\begin{equation}
G_\Phi(\w)=G_\Phi^0(\w)+\alpha[G_\Phi^0(\w)]^2 \chi_z(\w)
\label{exactrelation}
\end{equation}
where the bare bosonic Green's function $G_\Phi^0(\w)$ was introduced
in the previous equation~(\ref{GPhibare}). This connection can be easily
established from the use of equations of motions, or equivalently by using
standard manipulation of a path-integral representation of the problem.
The second step lies in a low-energy analysis of the bosonic propagator
$G_\Phi(\w)$ from the diagrammatic perspective. The key point is that
the bosonic self-energy $\Sigma_\Phi(\w)$, which was defined in Eq.~(\ref{fullPhi})
and expressed diagrammatically in Fig.~\ref{SigmaExact}, has a regular low-frequency 
behavior, owing to the role of the transverse magnetic field $\Delta$ in the
bare Majorana Green's function $G_y^0(\w)$, see Eq.~(\ref{barey}). As a
consequence, all bosonic self-energy diagrams are cut in the long-time limit
due to this energy gap, so that $\Sigma_\Phi(\w)=\Sigma_\Phi(0)+\mathcal{O}(\w)$
at low energy. Although the gap will be in reality filled by low energy excitations 
if one considers the full Majorana propagator $G_y(\w)$, we have seen in
Sec.~\ref{lowest} that a quickly vanishing tail of excitations is obtained, so that the 
above argument persists non-perturbatively (provided that perturbation theory is convergent). 
As a concrete illustration, equation~(\ref{SPhi1loop}) computed at one-loop
order demonstrates the regular character of the bosonic self-energy in the small
frequency limit. Accounting for the static part of the self-energy within the
explicit Dyson equation~(\ref{critical}), and neglecting the term
$\Sigma_\Phi(\w)-\Sigma_\Phi(0)$ at low energy, one obtains:
\begin{eqnarray}
G_\Phi(\w) \simeq \frac{1}{m_\Phi}- \frac{ib}{(m_\Phi)^2} \mathrm{Sign}(\w)|\w|^s
\end{eqnarray}
where the full mass $m_\Phi$ was given in Eq.~(\ref{mass}). Similarly, the bare
correlation function obeys a similar behavior:
\begin{eqnarray}
G_\Phi^0(\w) \simeq \frac{1}{m_\Phi^0}- \frac{ib}{(m_\Phi^0)^2} \mathrm{Sign}(\w)|\w|^s
\end{eqnarray}
now involving the bare mass $m_\Phi^0$.
We finally use the exact relation~(\ref{exactrelation}) in the low energy limit,
which imposes two conditions by taking the real and imaginary parts:
\begin{eqnarray}
\frac{1}{m_\Phi} &=& \frac{1}{m_\Phi^0}+\frac{\alpha}{(m_\Phi^0)^2}\chi_z'(0)\\
\nonumber
-\frac{b}{(m_\Phi)^2}\mathrm{Sign}(\w)|\w|^s  &=& 
-\frac{b}{(m_\Phi^0)^2}\mathrm{Sign}(\w)|\w|^s  
+\alpha \frac{\chi_z''(\w)}{(m_\Phi^0)^2} \\
&&\hspace{-0.0cm}  -\alpha \frac{2b}{(m_\Phi^0)^3}
 \chi_z'(0)\mathrm{Sign}(\w)|\w|^s
\end{eqnarray}
Solving this system to eliminate the mass $m_\Phi$ leads to the simple
expression:
\begin{eqnarray}
\chi_z''(\w)\simeq \frac{-b}{(m_\Phi^0)^2} \alpha |\w|^s 
[\chi_z'(0)]^2
\end{eqnarray}
asymptotically exact in the small $\w$ limit.
Using the known value for the coefficient $-b/(m_\Phi^0)^2=2\pi\w_c^{1-s}$ from 
Eq.~(\ref{ImGPhi}), and relation~(\ref{susceptibilities2}), we finally obtain the 
generalized Shiba relation Eq.~(\ref{shibasubohmic}).

\section{Quantum-to-classical mapping}

We can reformulate the usual quantum-to-classical mapping~\cite{Emery} in terms
of an effective quantum action in imaginary time calculated {\it perturbatively} in the coupling
constant $\alpha$.
Indeed, in the $\epsilon \rightarrow 0$ limit, we first notice that the linear form of the coupling
in the Hamiltonian Eq.~(\ref{ham}) leads to the exact relation:
\begin{equation}
\big< S^z \big> = \frac{s}{4\sqrt{\alpha}\w_c}\big<\Phi\big>
\end{equation}
which can be derived by the same strategy as discussed in Appendix~\ref{shiba}.
This result implies that the critical behavior of the impurity spin
magnetization $ \big< S^z \big>$ can be extracted from the knowledge of the bosonic
static average. The static critical properties can thus be alternatively understood from
the point of view of an effective action for the boson $\Phi$ only, which is
obtained by integrating out the spin degrees of freedom perturbatively in $\alpha$. 
This can be done diagrammatically using the Majorana fermions, leading after
some standard manipulations to the following expression in imaginary frequency:
\begin{equation}
S_\mr{eff}[\Phi] = \frac{1}{2} \int \frac{d\nu}{2\pi} \left(m_\Phi+A|\nu|^s\right) 
|\Phi(i\nu)|^2 + \int d\tau \; u \;[\Phi(\tau)]^4
\label{Seff}
\end{equation}
where $A\propto\w_c^{-1-s}$. The first important parameter here is the bosonic mass
$m_\Phi = s/(4\w_c) - \alpha/(2\Delta)+\mathcal{O}(\alpha^2)$, which agrees with
the one-loop result of Eq.~(\ref{SPhi1loop}), and controls the distance to
the quantum critical point at leading order. The second crucial term is the $\Phi^4$ 
interaction with coefficient $u=\alpha^2/(16\Delta^3)+\mathcal{O}(\alpha^3)$.

Within the {\it delocalized phase}, i.e.\ as long as the renormalized mass
$m_\Phi^R$ does not vanish, the bosonic spectral function obeys the low-frequency behavior 
$G_\Phi(i\nu) \propto 1/(m_\Phi^R+A|\nu|^s)
\propto 1/m_\Phi^R-A|\nu|^s/(m_\Phi^R)^2$, i.e.\ the slow decay in imaginary time
$G_\phi(\tau) \propto 1/\tau^{1+s}$. Owing to the exact
relation~(\ref{exactrelation}), we recover the correct small frequency behavior of
the longitudinal spin susceptibility, see Eq.~(\ref{GzG0W0}). The low-energy
sector of the spin-boson model is thus well captured by the effective bosonic
action~(\ref{Seff}), which turns out to be equivalent
to an Ising model in imaginary time with long-range spin exchange decaying as
$1/(\tau-\tau')^{1+s}$, exactly as expected from the quantum/classical
equivalence~\cite{Emery,Bulla}.

Let us finally discuss the low energy behavior at the quantum critical point.
Scaling analysis relative to the massless theory ($m_\Phi=0$) leads to the
bare scaling dimension $[u]=2s-1$. Thus, $u$ is relevant when $1/2<s$ and 
irrelevant otherwise. Therefore, in the weakly sub-ohmic regime $1/2<s<1$, 
the phase transition is described by an interacting fixed point~\cite{Fisher} 
and the magnetization $\big< S^z \big>$ obeys non-trivial classical exponents, 
as verified in the numerical study of Ref.~\onlinecite{Bulla}. On the contrary, in 
the strongly sub-ohmic range $0<s<1/2$, $u$ is irrelevant and mean-field 
theory should apply, a result that was recently 
debated.~\cite{Vojta,Kirchner,Rieger,Erratum,Anders,Feshke,Zhang,TongTrunc} The effective
action~(\ref{Seff}) would therefore tend to strengthen the idea that the
quantum-to-classical mapping is robust, unless the Landau parameter $u$ picks up
non-analyticities (such as discussed in Ref.~\onlinecite{BK1,BK2} for field theories
involving coupled fermions and bosons in higher dimensions). However, the
regularity of the diagrammatics would lead us to believe that this possibility
is unlikely for the spin-boson model.

\end{document}